\documentclass{aa}

\usepackage{amsmath,amssymb,amstext}
\usepackage{twoopt}
\usepackage[draft,breaklinks=true]{hyperref}
\usepackage{graphicx}
\usepackage{placeins}
\newcommand{\scoreExcellent}{\raisebox{-0.15ex}{\scalebox{1.30}{$\checkmark\!\checkmark$}}}
\newcommand{\scoreGood}{\raisebox{-0.15ex}{\scalebox{1.30}{$\checkmark$}}}
\newcommand{\scorePartial}{\raisebox{-0.15ex}{\scalebox{1.30}{$\sim$}}}
\newcommand{\scoreCalibrated}{\raisebox{-0.15ex}{\scalebox{1.30}{$\odot$}}}
\newcommand{\scoreFail}{\raisebox{-0.15ex}{\scalebox{1.30}{$\times$}}}

\begin{document}

\titlerunning{Lord/Egg of the LRDs}

\title{Lord/Egg of the LRDs: One geometry for its many faces}

\author{Piero Madau\inst{1,2}
\and
Roberto Maiolino\inst{3,4,5}
\and
Xihan Ji\inst{3,4}
\and
Xiaojing Lin\inst{6}
\and
Xiaohui Fan\inst{7}}

\institute{Dipartimento di Fisica ``G. Occhialini,'' Universit\`a degli
Studi di Milano-Bicocca, Piazza della Scienza 3, I-20126 Milano, Italy
\and
Department of Astronomy \& Astrophysics, University of California,
1156 High Street, Santa Cruz, CA 95064, USA
\and
Kavli Institute for Cosmology, University of Cambridge, Madingley Road,
Cambridge CB3 0HA, UK
\and
Cavendish Laboratory, University of Cambridge, 19 JJ Thomson Avenue,
Cambridge CB3 0HE, UK
\and
Department of Physics and Astronomy, University College London,
Gower Street, London WC1E 6BT, UK
\and
Department of Astronomy, Tsinghua University, Beijing 100084, China
\and
Steward Observatory, University of Arizona, 933 N. Cherry Ave.,
Tucson, AZ 85721, USA}

\abstract{Little red dots (LRDs) combine V-shaped ultraviolet--optical
continua, broad Balmer emission, and weak X-rays.
The nearby analog SDSS~J102530.29+140207.3 (known as ``Lord'' and ``the Egg'';
hereafter Lord/Egg) offers detailed spectroscopy and polarimetry. We test one
active galactic nucleus (AGN) geometry by fitting the ultraviolet-to-infrared
continuum with an inclination-dependent accretion-disk spectral energy
distribution (SED), a young stellar population, and a clumpy dusty screen. We use the
same SED in \textsc{Cloudy} calculations of a radially stratified virialized
broad-line region (BLR). The fit favors
$\log(M_{\rm BH}/M_\odot)\simeq7.4$, accretion at a highly
super-Eddington rate, $i\gtrsim70\degr$, and line-of-sight cloud extinction
$A_V^{\rm c}\sim3$~mag, and predicts a 2-keV
luminosity below the \textit{Chandra} limit. A BLR whose density rises from
$n_{\rm H}=10^9$ to $10^{11}\,{\rm cm}^{-3}$ near the sublimation radius
reproduces the broad H$\alpha$ emission profile without requiring a separate intermediate-width
Gaussian component. For cloud columns of order
$N_{\rm H}^{\rm c}\sim10^{23}\,{\rm cm}^{-2}$, it also matches the different
widths and exponential-like broad wings of the Balmer lines without requiring
electron scattering, and reproduces the Balmer decrement. The redshifted H$\beta$--H$\delta$ absorber lacks
an H$\alpha$ counterpart; the nearly line-independent predicted covering
therefore requires velocity-coherent H$\alpha$ re-emission into the absorption
trough. Cool outer clumps
account for K~{\sc i} and saturated Na~{\sc i} absorption, while the geometry
recovers the continuum and H$\alpha$ polarization levels and their relative
position-angle offset. Ca~{\sc ii}
triplet absorption requires localized, non-radiatively heated, warm dust-poor
zones near the sublimation front. He~{\sc i} absorption traces a fast
($\simeq500\,\mathrm{km\,s^{-1}}$), low-column ionized BLR skin. Together
with slower blueshifted low-ionization gas and red Balmer absorption, all
projected absorber velocities remain below the characteristic local escape
speeds, consistent with a
bound failed BLR wind. Therefore, a dusty, high-inclination accretion
system can explain Lord/Egg within one connected geometry.}


\maketitle

%=====================================================================
\section{Introduction}
\label{sec:intro}

The \textit{James Webb Space Telescope} has uncovered many compact,
high-redshift broad-line sources with red rest-optical colors, blue
ultraviolet continua, large Balmer equivalent widths (EWs), and weak X-rays.
These ``little red dots'' (LRDs) may signal rapid
black-hole growth, but differ from ordinary unobscured quasars
\citep[e.g.][]{Greene2024,Matthee2024,Juod2024JADES,Kocevski2025,Geris2026}.
Proposed explanations
include reddened active galactic nuclei (AGNs), super-Eddington flows, and
dense reprocessing envelopes, usually tested against only part of the
available data.

The nearby source SDSS~J102530.29+140207.3 ($z=0.1006$; J1025) provides a
stringent test. Initially identified as a metal-poor dwarf with persistent
broad H$\alpha$ \citep{IzotovThuan2008}, it is X-ray underluminous and shows a
narrow blueshifted H$\alpha$ absorber against its broad-line region (BLR)
\citep{Simmonds2016,Burke2021}. \citet{Ji2025z01} named it ``Lord,'' and
\citet{Lin2026} called it ``the Egg''; we use Lord/Egg. Its V-shaped
continuum, broad Balmer and Paschen lines, Balmer and He~{\sc i} absorption,
exceptionally strong Na~{\sc i}, K~{\sc i}, and Ca~{\sc ii} absorption,
weak X-rays, and mid-infrared emission closely reproduce the LRD
phenomenology. The smooth turnover near the Balmer limit shows no secure
discontinuity and is interpreted here as the crossover between young stars
and a reddened disk.

Alternative interpretations remain viable. \citet{Lin2026} proposed a
largely thermalized, $T\sim5000$~K envelope; atmosphere calculations reproduce
the optical--near-infrared continuum and Ca~{\sc ii} triplet (CaT) absorption,
but underpredict Na~{\sc i}~D \citep{Liu2026Atmospheres}. Other diagnostics
favor structured, anisotropic gas. Strong Balmer absorption is rare among
low-redshift type-1 AGNs and requires optically thick H$\alpha$ absorption
with projected covering above about 0.8 \citep{Shangguan2026}. Extreme LRD
Balmer decrements also depend on dense-gas transfer and which cloud faces are
visible, but the most extreme ratios still require dust reddening
\citep{Kiyota2026}.
VLT/FORS2 data further show nearly gray continuum polarization, weaker broad
H$\alpha$ polarization, and a $48\degr$ offset between the continuum and
broad-H$\alpha$ polarization angles
\citep{DEugenio2026Polarization}, reproduced by combining intrinsic disk
polarization with dichroic transmission through misaligned dust
\citep{MadauMaiolino_POL_2026}.

We test the picture in which LRDs are dust-intercepted, high-inclination
counterparts of compact blue broad-line AGNs powered by super-Eddington
accretion \citep{MadauMaiolino2026}. A thick flow supplies an
angle-dependent ionizing spectrum, while a radially stratified virial BLR and
an outer clumpy dusty medium analogous to the obscuring structure in the
standard AGN unified model set the line and continuum properties.
Emissivity-weighted radii
naturally produce nearly exponential wings \citep{Scholtz2026} and predict a
larger H$\alpha$ than H$\beta$ half-light radius
\citep{Bentz2010,Feng2021}; a common electron-scattering layer cannot produce
line-dependent widths \citep{Torralba2026b}. We combine this spectrum with
young stars, dust, and photoionization calculations. The continuum fixes the
central emission and attenuation, H$\alpha$ fixes the BLR covering and radial
velocity field, and the remaining lines, infrared emission, and polarization
provide independent tests.

Throughout, $C_i$ is the sky covering fraction of component $i$ as seen
from the source. We use $C_{\rm dust}$ for the total dust covering,
$f_{\rm hd}\equiv L_{\rm hd}/L_{\rm disk}^{\rm bol}$ for the hot/warm
reradiated fraction, and $C_{{\rm los},j}$ for the flux-weighted fraction of
the background source covered by absorber $j$. For optically thick clouds,
$C_{\rm dust}\gtrsim f_{\rm hd}$ if colder reradiation is unobserved.

%=====================================================================
\section{The ultraviolet-to-near-infrared continuum}
\label{sec:sed}

\subsection{Data and forward model}
\label{sec:sed_model}

We fitted the 11 GALEX, SDSS, UKIDSS, and WISE bands at rest-frame
$\lambda<2.5\,\mu$m from the compilation of \citet{Ji2025z01}; longer
wavelengths, where dust reradiation matters, were fitted subsequently. The
model was integrated through each filter, with published line fluxes added
to the synthetic photometry. This correction is largest in the SDSS $i$
band containing H$\alpha$ and remains non-negligible elsewhere
\citep{Ji2025z01,Lin2026}. Direct electron-temperature measurements give
$12+\log({\rm O/H})=7.58\pm0.03$ \citep{Lin2026} and
$7.73^{+0.21}_{-0.14}$ \citep{Ji2025z01}, or
$0.08$--$0.11\,Z_\odot$; this gas abundance is distinct from the stellar
metallicity below.

The nuclear component is the angle-dependent spectral energy distribution
(SED) of the thick-plus-thin disk of \citet{Madau2026}, applied to
LRD-to-little-blue-dot (LBD) unification by
\citet{MadauMaiolino2026}, interpolated in $M_{\rm BH}$,
$\dot m\equiv0.1\dot M c^2/L_{\rm Edd}$, and inclination $i$. We attenuated
it with a smooth, bump-free dust extinction curve based on a modification of 
the \citet{Cardelli1989} law,
fitting $A_V^{\rm c}$ and $R_V$: over $2500$--$8300$~\AA,
$\ln(A_\lambda/A_V)$ is a cubic in $\ln\lambda$ joined to the Cardelli et al. ultraviolet
and infrared branches. The second component is an unreddened stellar
population with a constant star formation rate (SFR), a Chabrier initial mass
function, and fixed metallicity
$Z_\star=0.2\,Z_\odot$ \citep{BruzualCharlot2003,Chabrier2003}. The stellar
template omits nebular continuum; measured emission-line fluxes are added
separately to the synthetic photometry. Its intrinsic
far-ultraviolet slope is $\beta_{\rm FUV}=-3.1$, where
$F_\lambda\propto\lambda^{\beta_{\rm FUV}}$ over $1300$--$2500$~\AA.
Extended ultraviolet morphologies in some LRDs support such a stellar
contribution, although Lord/Egg is unresolved \citep{Ji2025z01,Lin2026}.
Young stars can then supply the blue ultraviolet light while the compact
disk is strongly reddened by a steep law; assigning both SED branches to the
disk would instead require grayer attenuation, as in
\citet{MadauMaiolino2026}. We added no independent
Balmer-continuum opacity because no unambiguous discontinuity is observed.

Schematically, before convolution with the filters, the model is
\begingroup
\setlength{\abovedisplayskip}{4pt}
\setlength{\belowdisplayskip}{4pt}
\setlength{\abovedisplayshortskip}{4pt}
\setlength{\belowdisplayshortskip}{4pt}

%---------------------------------------------------------------------
\begin{equation}
 F_{\lambda_{\rm obs}}=
 \frac{L_\lambda^{\rm disk}10^{-0.4A_\lambda}+L_\lambda^\star}
 {4\pi D_L^2(1+z)}+F_{\lambda_{\rm obs}}^{\rm lines},
 \quad \lambda_{\rm obs}=(1+z)\lambda.
 \label{eq:sed_model}
\end{equation}
%---------------------------------------------------------------------
\endgroup
Here, $D_L$ and $z=0.1006$ are the luminosity distance and redshift;
$L_\lambda^{\rm disk}$ is the rest-frame, angle-dependent disk luminosity;
and $A_\lambda$ is set by $A_V^{\rm c}$ and $R_V$. In the minimal interpretation,
$A_V^{\rm c}$ is the extinction produced by a single intercepted cloud. If
multiple identical clouds are intercepted, the fit instead constrains their
summed extinction. The stellar
luminosity $L_\lambda^\star$ is set by its age $t_\star$ and SFR and remains
unreddened because the extended population need not share the nuclear
sightline. The measured, already attenuated line contribution
$F_{\lambda_{\rm obs}}^{\rm lines}$ enters the photometry, whereas
\textsc{Cloudy} line predictions enter only the BLR analysis.

We fitted $(M_{\rm BH},\dot m,i,A_V^{\rm c},R_V,t_\star,{\rm SFR})$, adopting
weak priors $\log(M_{\rm BH}/M_\odot)=7.0\pm0.5$ and
$i=75^\circ\pm10^\circ$, together with the bounds $2.2\leq R_V\leq6$, and
$t_\star\geq3$~Myr. The mass prior spans the single-epoch estimates and their
approximately 0.5 dex systematic uncertainty. Because inclination is
degenerate with $\dot m$, mass, and extinction, its prior encodes our 
high-inclination interpretation rather than an independent measurement.
The Balmer decrement was reserved as a subsequent BLR test.

\begin{figure}
\centering
\includegraphics[width=0.99\columnwidth]{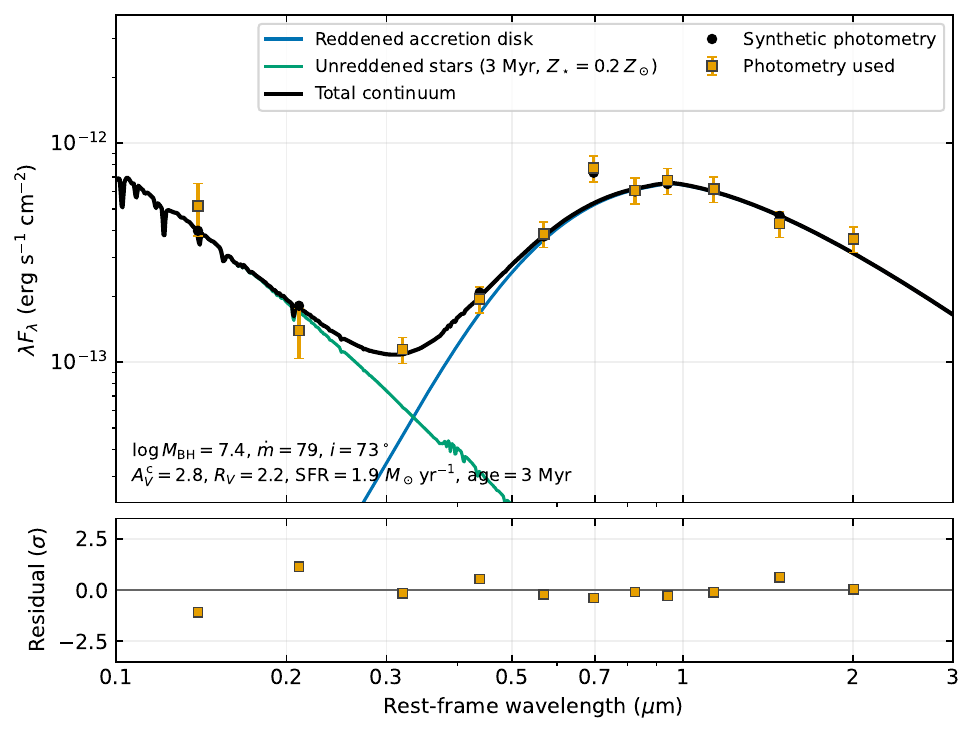}
\caption{Ultraviolet-to-near-infrared fit to the photometry compiled by
\citet{Ji2025z01}, including measured line contamination
\citep{Ji2025z01,Lin2026}. Orange squares are fitted data, black circles are
synthetic photometry, and the curves show only the continuum components:
the reddened disk (blue), stars (green), and their sum (black). The lower panel
gives logarithmic residuals.
The fit has $\chi_\nu^2=0.9$ and uses neither dust reradiation nor an arbitrary
disk normalization.}
\label{fig:sed}
\end{figure}

\subsection{Best-fitting continuum}
\label{sec:sed_results}

Figure~\ref{fig:sed} shows the resulting decomposition. The best-fitting
parameters are
%---------------------------------------------------------------------
\begin{align}
 \log(M_{\rm BH}/M_\odot)&=7.4, & \dot m&=80,\nonumber\\
 i&=73^\circ, & A_V^{\rm c}&=2.8~{\rm mag},\nonumber\\
 R_V&=2.2, & t_\star&=3~{\rm Myr},\nonumber\\
 {\rm SFR}&=1.9~M_\odot\,{\rm yr}^{-1}.&&
 \label{eq:sed_best}
\end{align}
%---------------------------------------------------------------------
%
The V-shaped SED is the crossover of far-ultraviolet starlight and the
reddened disk, which dominates beyond the near-ultraviolet minimum. AGN-only
trials failed even with gray attenuation, whereas the young population fits
the GALEX point while contributing only about 2\% of the disk continuum at
H$\alpha$. Its boundary age of 3~Myr describes the UV-bright episode, not a
precise host age; older templates add unneeded optical light and a
Balmer/4000~\AA\ discontinuity. Because $R_V$ and $\dot m$ reach grid
boundaries, they indicate steep attenuation and rapid accretion rather than
precise values. The fitted $\dot m=80$ corresponds to
$\dot M\simeq45\,M_\odot\,\mathrm{yr^{-1}}$, about 25 times the SFR, and a
nominal $M_{\rm BH}/\dot M\simeq0.6$~Myr growth time; outflow losses would
lengthen the actual e-folding time. Lord/Egg may therefore be in a brief,
intensely fueled phase whose gas supply remains unidentified.
Comparably low \(R_V\) values have indeed been reported: reddened quasars span $R_V=2.2$--2.7, with
an ensemble value $2.21\pm0.22$ \citep{Zafar2015}. Our AGN-only LRD models
instead required $R_V\simeq4$--5.5 to retain blue nuclear ultraviolet light
\citep{MadauMaiolino2026,MadauMaiolinoLF2026}; here, stars supply that light
and allow steeper nuclear attenuation. Nonstandard extinction curves have
also been inferred at high redshift, including in a $z=4.41$ galaxy, a
$z=4.6$ QSO, and dusty circumgalactic sightlines
\citep{Sanders2025Attenuation,Li2026SteepExtinction,Sun2026SmallGrains}.

The fitted mass provides an independent consistency check. Based on
single-epoch virial relations, \citet{Ji2025z01} obtained
$\log(M_{\rm BH}/M_\odot)=6.49\pm0.01$ without attenuation and
$7.22\pm0.04$ with their maximum Balmer-decrement correction;
\citet{Lin2026} found $6.52\pm0.03$, with dust potentially adding
0.3--0.7 dex. Our continuum-based value of 7.4 agrees with the corrected estimates
within the approximately 0.5 dex calibration uncertainty, but exceeds the
uncorrected values by nearly 0.9 dex. 

%=====================================================================
\section{Dust reradiation in the infrared}
\label{sec:infrared}

\begin{figure}
\centering
\includegraphics[width=0.99\columnwidth]{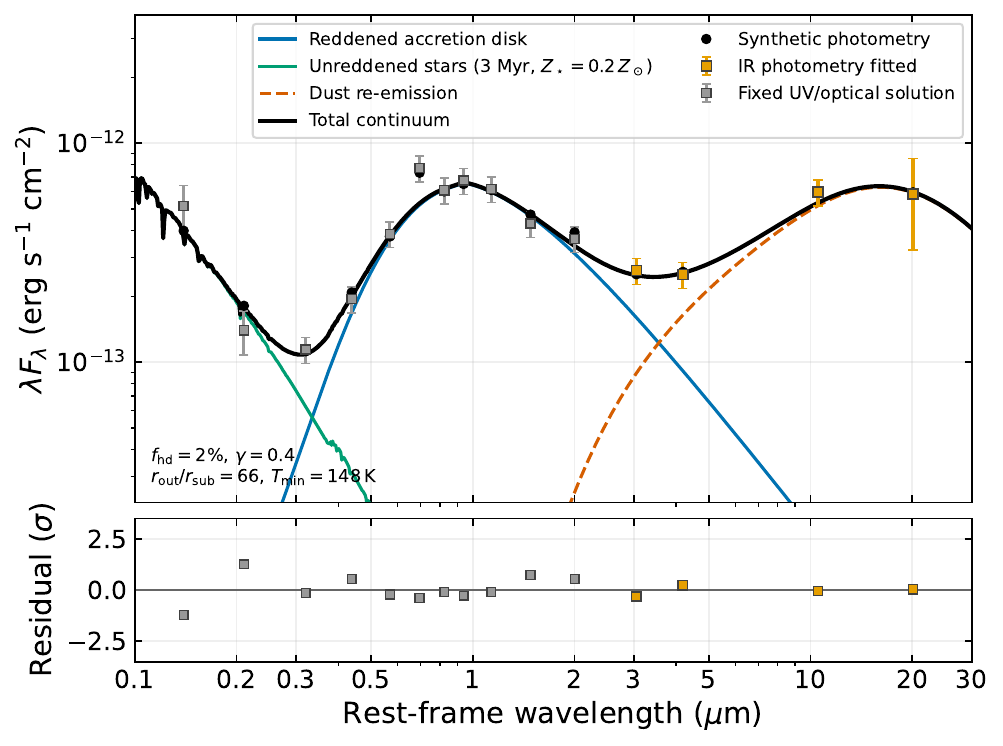}
\caption{Infrared fit to the photometry of \citet{Ji2025z01}, with the
disk-plus-stellar solution fixed. Gray squares are the shorter-wavelength
data and orange squares the fitted infrared measurements. Solid curves show
the attenuated disk continuum (blue), stellar continuum (green), and total
continuum (black); the dashed orange curve shows dust reradiation. The
conditional fit has $\chi_\nu^2=0.2$; colder dust beyond the
observed range remains unconstrained. The fitted $f_{\rm hd}$ is the
hot/warm reradiated fraction, not the total dust covering.}
\label{fig:infrared}
\end{figure}

Holding the disk and stellar solution fixed, we fitted the four bands beyond
rest-frame $2.5\,\mu$m. Their coverage ends near $20\,\mu$m and therefore
constrains warm dust, not the colder far-infrared peak. We used gray thermal
spectra extending from a $1200$~K sublimation front, with
$T\propto r^{-1/2}$ as expected for gray radiative equilibrium. We fitted the
hot/warm reradiated fraction $f_{\rm hd}$, the dust-number-density slope
$\gamma$ defined by $n_{\rm dust}(r)\propto r^{-\gamma}$, and
$r_{\rm out}/r_{\rm sub}$. The luminosity fraction $f_{\rm hd}$ is
distinct from both the total dust covering $C_{\rm dust}$ and the extinction
$A_V^{\rm c}$ along our particular sightline.
Figure~\ref{fig:infrared} shows the resulting fit. The best-fitting
parameters are
%---------------------------------------------------------------------
\begin{equation}
 f_{\rm hd}=0.02,\qquad \gamma=0.4,\qquad
 r_{\rm out}/r_{\rm sub}=66.
 \label{eq:infrared_best}
\end{equation}
%---------------------------------------------------------------------
The outer radius corresponds to about $150$~K. With four data points and
three fitted parameters, $\chi_\nu^2=0.2$ does not uniquely determine the
radial distribution. Because the data end at rest-frame $20\,\mu$m,
$f_{\rm hd}=0.02$ measures only the hot/warm nuclear component; cooler
AGN-heated dust can make $C_{\rm dust}>f_{\rm hd}$, while colder star-heated
host dust is unconstrained. The small $f_{\rm hd}$ can coexist with large
line-of-sight extinction along a cloud-intercepting sightline. The
population-wide $\langle C_{\rm dust}\rangle\simeq0.2$
\citep{MadauMaiolinoLF2026} is therefore not directly comparable.

Integrating the observed $0.3$--$3\,\mu$m bump and modeled $1$--$100\,\mu$m
dust component gives $L_{\rm hd}/L_{\rm opt,obs}\simeq1.3$. The observed
optical bump therefore cannot be the intrinsic heating continuum: the dust
intercepts a substantially brighter ultraviolet--optical source that is
strongly attenuated and self-occulted along our sightline.

%=====================================================================
\section{X-ray weakness and viewing angle}
\label{sec:xray}

Two physically distinct X-ray components are relevant. Comptonization in
a compact corona produces non-thermal hard X-rays, whereas the optically thick
disk atmosphere contributes a thermal soft-X-ray Wien tail. Our continuum
model predicts only the latter.

Lord/Egg was not detected in either the archival \textit{Chandra} observation or
the deeper \textit{NuSTAR} data. The latter primarily constrains a harder
or more heavily obscured coronal component, whereas \textit{Chandra} gives the
tighter constraint below 10~keV. \citet{Ji2025z01} derived a 90\% confidence
upper limit
$L_{2-10\,\mathrm{keV}}<2.8\times10^{41}\ \mathrm{erg\,s^{-1}}$ from
\textit{Chandra}. All X-ray energies quoted below are in the rest frame.
This corresponds to the monochromatic limit
%
%---------------------------------------------------------------------
\begin{equation}
 \nu L_\nu(2\,\mathrm{keV})<
 1.4\times10^{41}\ \mathrm{erg\,s^{-1}}.
 \label{eq:xray_limit}
\end{equation}
%---------------------------------------------------------------------
%
We do not calculate the absolute coronal luminosity, which requires a
detailed prescription for the coronal energy budget and radiative transfer.
\citet{Madau2024} showed that intense illumination by the surrounding
funnel walls enhances Compton cooling. Building on this mechanism,
\citet{Trinca2026} obtained $kT_e\lesssim40$~keV and
$\Gamma\simeq2.8$--4 for a corona confined and radiatively over-cooled within
a narrow super-Eddington funnel (their Fig.~1), naturally accounting for
systematic \textit{Chandra} nondetections of high-redshift broad-line AGNs.
Steep X-ray spectra consistent with cold coronae are also widespread among
luminous $z\gtrsim6$ quasars \citep{Tortosa2026}. At Lord/Egg's inclination,
direct coronal emission is additionally occulted by the optically thick disk.

We compute the thermal component by extending the same angle-dependent
disk-atmosphere model used in the continuum fit to 2~keV. The maximum surface
temperature is $T_{\rm max}\simeq2.5\times10^6$~K
($kT_{\rm max}\simeq0.21$~keV) near $2.7r_{\rm S}$, where
$r_{\rm S}=2GM_{\rm BH}/c^2$. These innermost annuli dominate the directly
visible Wien tail in near-polar views. At $i=73\degr$ they are self-occulted;
the residual high-inclination component is radiation returned by the visible
funnel walls in the frequency-conserving self-irradiation calculation. For
the best-fitting values $\log(M_{\rm BH}/M_\odot)=7.4$ and $\dot m=80$, the
unabsorbed disk-atmosphere component is
%---------------------------------------------------------------------
\begin{equation}
 \nu L_\nu(2\,\mathrm{keV})=
 \begin{cases}
  1.2\times10^{44}\ \mathrm{erg\,s^{-1}}, & i=20^\circ,\\
  8\times10^{40}\ \mathrm{erg\,s^{-1}}, & i=73^\circ.
 \end{cases}
 \label{eq:xray_inclination}
\end{equation}
%---------------------------------------------------------------------
Here $i=73^\circ$ is the best-fitting inclination in
Eq.~(\ref{eq:sed_best}), while $i=20^\circ$ illustrates a near-polar
view. The corresponding intrinsic disk values of
$\nu L_\nu(2500\,\text{\AA})$ are
$1.7\times10^{45}$ and $4.1\times10^{44}\ \mathrm{erg\,s^{-1}}$,
respectively. Defining
%---------------------------------------------------------------------
\begin{equation}
 \alpha_{\rm OX}=0.3838\log\!\left[
 \frac{L_\nu(2\,\mathrm{keV})}{L_\nu(2500\,\text{\AA})}\right],
 \label{eq:alphaox}
\end{equation}
%---------------------------------------------------------------------
we obtain $\alpha_{\rm OX}=-2.4$ at $73^\circ$ and $-1.5$ at
$20^\circ$. At the corresponding near-polar UV luminosity, the empirical
$\alpha_{\rm OX}$--$L_\nu(2500\,\text{\AA})$ relation for unobscured
quasars predicts $\alpha_{\rm OX}\simeq-1.5$ \citep{Just2007}. Thus the
near-polar thermal disk contribution alone reaches a comparatively normal
X-ray-to-UV ratio. The factor of approximately $1.5\times10^3$ suppression at
high inclination is much stronger than the UV anisotropy and produces
apparent X-ray weakness through self-occultation.

Photoelectric absorption along the line of sight will further attenuate
the high-inclination funnel-wall component and any residual coronal emission.
For one of the dust-free BLR clouds adopted below,
%---------------------------------------------------------------------
\begin{equation}
 \tau_{\rm pe}(E)=N_{\rm H}^{\rm c}\,\sigma_{\rm pe}(E,Z,U),
 \qquad
 \tau_{\rm T}=N_{\rm H}^{\rm c}\sigma_{\rm T},
 \label{eq:xray_tau}
\end{equation}
%---------------------------------------------------------------------
where $N_{\rm H}^{\rm c}=10^{23}\,\mathrm{cm}^{-2}$ is the column per
cloud, $\sigma_{\rm pe}$ is the photoelectric absorption cross section per H
nucleus, and
$\sigma_{\rm T}=6.65\times10^{-25}\,\mathrm{cm}^{2}$ is the Thomson
scattering cross section. The resulting Thomson depth is only $\tau_{\rm T}=0.07$, so
multiple-scattering corrections to the photon path length and absorption
probability are small. For neutral gas, the photoelectric cross section at
2~keV is approximately $(3$--$4)\times10^{-23}\,\mathrm{cm}^{2}$ per H
nucleus at solar abundance \citep{Wilms2000}. At the adopted $0.1\,Z_\odot$,
the H and He opacity prevents a simple linear scaling with metallicity; we
estimate $\sigma_{\rm pe}(2\,\mathrm{keV})\simeq
(6$--$8)\times10^{-24}\,\mathrm{cm}^{2}$, or
$\tau_{\rm pe}\simeq0.6$--0.8. The unscattered direct fraction,
$\exp[-(\tau_{\rm pe}+\tau_{\rm T})]$, is roughly 40--50\% at 2~keV.
Applying this attenuation to the high-inclination funnel-wall
component gives
$\nu L_{\nu,{\rm obs}}^{\rm th}(2\,\mathrm{keV})\simeq
(3.4$--$4.1)\times10^{40}\,\mathrm{erg\,s^{-1}}$, below the
\textit{Chandra} limit. This estimate assumes the adopted $0.1\,Z_\odot$
abundance. The cloud is effectively opaque below
approximately 1~keV.

\begin{figure}[t]
\centering
\includegraphics[width=0.99\columnwidth]{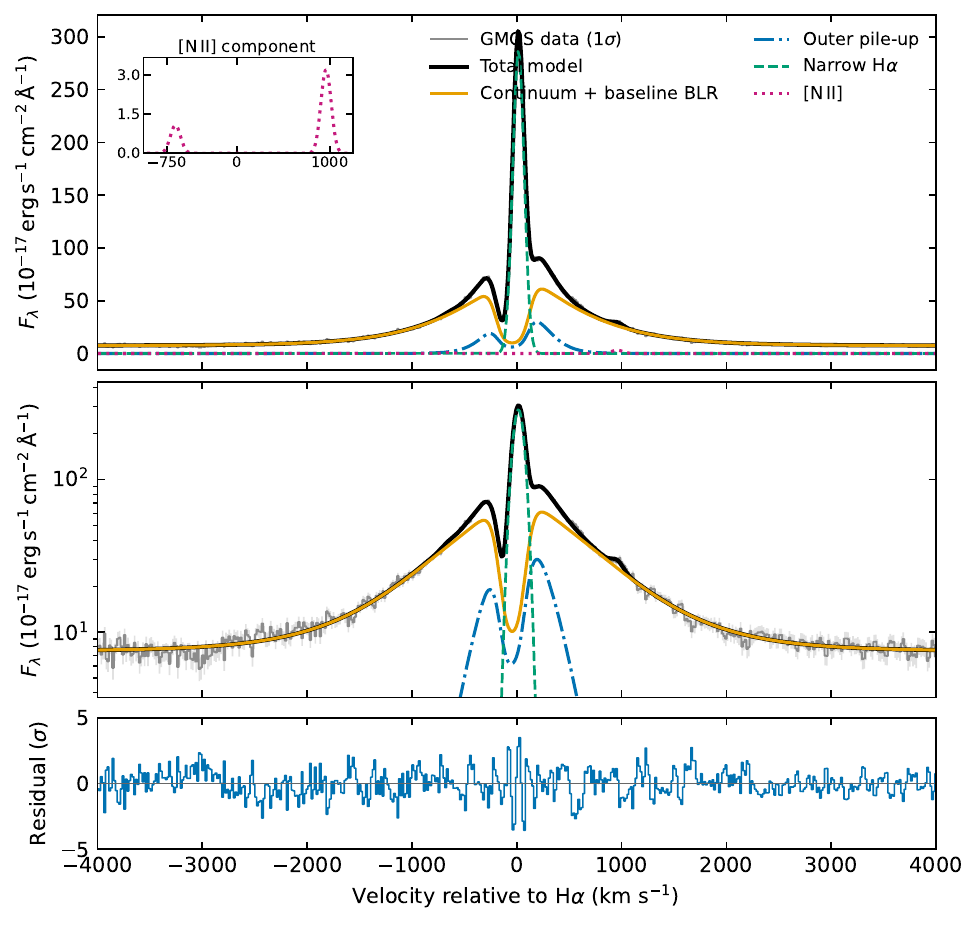}
\caption{Stratified-BLR fit to the Gemini/GMOS H$\alpha$ spectrum
\citep{Burke2021,Ji2025z01}. For $f_{\rm vir}=1$, the radial model gives
$p=1.55$ and $\chi_\nu^2=1.07$. Curves show the total (black), continuum plus
baseline BLR (orange), dense $r_{\rm sub}$ pile-up (blue dot-dashed), narrow
H$\alpha$ (green dashed), and [N~{\sc ii}] (magenta dotted; enlarged in the
inset). The blue absorption trough is included; the red trough is almost
exactly refilled (Sect.~\ref{sec:missing_halpha}). The middle and lower panels
show the profile logarithmically and the residuals.}
\label{fig:halpha_profile}
\end{figure}

%=====================================================================
\section{Broad-line emission and BLR stratification}
\label{sec:halpha}

\subsection{A photoionized, radially stratified BLR}
\label{sec:blr_model}

We next used the same angle-dependent disk SED and \textsc{Cloudy}
calculations to model both the luminosity and the shape of the broad
H$\alpha$ line. In a radially stratified BLR, clouds spanning a range of
radii experience different ionizing fluxes and have different virial
velocities. Their line weights are therefore not adjusted independently:
the H$\alpha$ contribution at each radius follows from the incident SED,
the photoionized gas emissivity, and the adopted cloud distribution
\citep{Madau2026Wings}.

We computed the line emissivities with the C23 release of \textsc{Cloudy}
\citep{Chatzikos2023}. The incident radiation was drawn from the same
family of angle-dependent thick-plus-thin disk SEDs used for the continuum
fit. To represent the flattened BLR, we sampled cloud illumination angles
$i_{\rm BLR}=65^\circ,70^\circ,75^\circ,80^\circ$, and $85^\circ$, measured
from the disk axis and distinct from the observer inclination. We combined
their contributions using the equatorially concentrated angular cloud
distribution of \citet{MadauMaiolino2026}, proportional to
$\exp[-\cos^2(i_{\rm BLR})/(2\sigma_{\rm BLR}^2)]$, with dimensionless angular
width $\sigma_{\rm BLR}=0.2$.
For each illumination angle, the $n_{\rm H}=10^9\,{\rm cm}^{-3}$ grid
covered ionization parameters $-4\leq\log U\leq1.5$; matched grids at $10^{10}$ and
$10^{11}\,{\rm cm}^{-3}$ sampled the outer-BLR range used below. Each model
represented a dust-free, constant-density, plane-parallel BLR cloud with
$N_{\rm H}^{\rm c}=10^{23}\,{\rm cm}^{-2}$ and metallicity
$Z=0.1\,Z_\odot$.
For a cloud illuminated at polar angle $i_{\rm BLR}$,
%---------------------------------------------------------------------
\begin{equation}
 U(r,i_{\rm BLR})=
 \frac{Q_{\rm H}(i_{\rm BLR})}
 {4\pi r^2 n_{\rm H}c},
 \label{eq:u_radius}
\end{equation}
%---------------------------------------------------------------------
where $Q_{\rm H}(i_{\rm BLR})$ is the angle-dependent hydrogen-ionizing
photon rate of the thick-plus-thin disk and $U$ is the dimensionless ratio of
the hydrogen-ionizing photon density to the hydrogen density. Each point in the \textsc{Cloudy}
grid therefore maps to a physical radius while retaining the angular
variation of the incident continuum and the nonlinear dependence of the line
emissivity on $U$.

For the continuum solution in Eq.~(\ref{eq:sed_best}) and
$T_{\rm sub}=1200$~K, we obtained $r_{\rm sub}=1.7$~pc. We imposed an inner
boundary at $r_{\rm min}=0.01\,r_{\rm sub}$.
We parameterized the radial distribution directly through the
differential BLR covering factor, adopting a power law with an exponential
outer taper and adding a localized enhancement at the BLR--dust transition,
%---------------------------------------------------------------------
\begin{equation}
 \frac{dC_{\rm BLR}}{dr}
 = {\cal N}\left[
 x^{-p}\exp(-x)+A_{\rm sub}
 \exp\!\left(-\frac{(x-1)^2}{2s_{\rm sub}^2}\right)
 \right],
 \label{eq:blr_radial}
\end{equation}
%---------------------------------------------------------------------
%
where $x\equiv r/r_{\rm sub}$ and ${\cal N}$ normalizes the integrated
covering to $C_{\rm BLR}$. The first term describes the baseline BLR; the
second is a pile-up centered at $r_{\rm sub}$, with relative amplitude
$A_{\rm sub}$ and fractional width $s_{\rm sub}$. The baseline term combines
the power-law covering used in global BLR models \citep{Netzer2020} with the
outer taper adopted by
\citet{Madau2026Wings}.\footnote{\citet{Madau2026Wings} parameterized the
shell-integrated cloud number as $dN/dr\propto
(r/r_{\rm sub})^{\alpha-1}\exp(-r/r_{\rm sub})$ and assumed a cloud area
independent of radius. Here, $A_c$ is the projected physical area of an
individual BLR cloud. Since $dC_{\rm BLR}=A_c\,dN/(4\pi r^2)$, the two
notations are equivalent, with $p=3-\alpha$. More generally, if
$A_c\propto r^q$, only the combination $p=3-\alpha-q$ is constrained by the
mean spectrum and smooth line profiles.}
The pile-up is motivated by failed dusty-wind and dust-inflated-disk models,
which concentrate lifted or returning material near the sublimation boundary
\citep{CzernyHryniewicz2011,BaskinLaor2018,NaddafCzerny2022,
NaddafCzernyZajacek2022}; its amplitude and extent remain phenomenological.
Its velocity width follows from its radius: at $r_{\rm sub}=1.7\,{\rm pc}$,
Eq.~(\ref{eq:blr_velocity}) gives $\sigma_v\simeq250\,{\rm km\,s^{-1}}$, or
${\rm FWHM}\simeq590\,{\rm km\,s^{-1}}$.
We separately interpolate the density from the effective inner-BLR value
$n_{\rm H}=10^9\,{\rm cm}^{-3}$ to $10^{11}\,{\rm cm}^{-3}$ near
$r_{\rm sub}$, using matched \textsc{Cloudy} grids at
$\log_{10}(n_{\rm H}/{\rm cm}^{-3})=9$, 10, and 11. Dense gas at the
sublimation boundary has previously been invoked to explain
intermediate-width permitted emission in AGNs
\citep{Adhikari2016,Adhikari2017}, while
$n_{\rm H}\sim10^{11.5}$--$10^{12}\,{\rm cm}^{-3}$ has been inferred for
outer low-ionization O~{\sc i}-, Ca~{\sc ii}-, and Fe~{\sc ii}-emitting
quasar gas \citep{Matsuoka2007}. Here the density increase represents
compressed disk-atmosphere or stalled failed-wind gas.

The exponent $p$ controls the baseline component's radial concentration, with
larger values weighting the inner BLR more strongly.
At a given radius, the line-of-sight cloud velocities were represented by
a Gaussian of dispersion
%
%---------------------------------------------------------------------
\begin{equation}
 \sigma_v(r)=f_{\rm vir}\left(\frac{GM_{\rm BH}}{r}\right)^{1/2},
 \label{eq:blr_velocity}
\end{equation}
%---------------------------------------------------------------------
%
where $f_{\rm vir}$ is an effective one-dimensional velocity-scale parameter,
rather than an inclination factor alone; it absorbs projected rotation,
radial and vertical motions, and cloud-to-cloud velocity dispersion.
Eq.~(\ref{eq:blr_velocity}) gives the Gaussian kernel at one radius, but
the observed FWHM must be measured from the luminosity-weighted profile
integrated over all radii. The emergent profile is then
%---------------------------------------------------------------------
\begin{equation}
 \begin{split}
 \frac{dL_{\rm H\alpha}}{dv}={}&
 \int dr\,4\pi r^2
 \int di_{\rm BLR}\,p(i_{\rm BLR})\\
 &\times \frac{dC_{\rm BLR}}{dr}
 F^c_{\rm H\alpha}[U(r,i_{\rm BLR}),n_{\rm H}(r)]
 G[v;\sigma_v(r)],
 \end{split}
 \label{eq:halpha_profile}
\end{equation}
%---------------------------------------------------------------------
%
where $F^c_{\rm H\alpha}$ is the emergent \textsc{Cloudy} line
surface flux interpolated in $U$ and $n_{\rm H}$, and $p(i_{\rm BLR})$ is the
normalized angular cloud distribution. The more distant dusty portions of
the dense outer population enter the absorption calculations but contribute
negligibly to this permitted-line emission integral.
The grid uses these emergent surface fluxes to construct cloud
luminosities in the radial integral; it does not explicitly ray-trace whether
the observer sees the illuminated face, the shielded face, or a mixture of
the two. The inner clouds generate the high-velocity tails, whereas larger radii contribute
preferentially near the line core. This luminosity-weighted superposition of
Gaussian velocity kernels can produce nearly exponential wings, with their
relative amplitudes determined self-consistently by the radial covering
distribution and \textsc{Cloudy} emissivity.

\subsection{H$\alpha$ emission profile and Paschen diagnostics}
\label{sec:halpha_results}

Because MODS-R provides substantially higher signal-to-noise but GMOS
provides higher spectral resolution, we used the MODS-R spectrum to set the
absolute broad H$\alpha$ normalization and the GMOS spectrum to constrain its
emission-line profile shape. Our decomposition of MODS-R gives a rest-frame
broad H$\alpha$ emission equivalent width
$W_{\rm H\alpha}=357\pm18$~\AA. We normalized the complete
covering distribution, including its pile-up near $r_{\rm sub}$, to this
flux. Iterating the angular weights, radial distribution, and H$\alpha$
normalization, while including the small stellar continuum contribution at
H$\alpha$, gave
%---------------------------------------------------------------------
\begin{equation}
 C_{\rm BLR}=0.02.
 \label{eq:cblr}
\end{equation}
%---------------------------------------------------------------------
This is numerically equal to the separately inferred hot/warm reradiated
fraction, $f_{\rm hd}=0.02$ (Sect.~\ref{sec:infrared}), and is suggestive of a
low-covering, clumpy structure extending across the BLR--dust sublimation
transition. However, the agreement should not be overinterpreted:
$C_{\rm BLR}$ normalizes the dust-free line-emitting clouds, whereas
$f_{\rm hd}$ is an energy-budget measure of the modeled dust component, and
both values depend on the adopted disk luminosity.
For the profile shape, we fitted the Gemini/GMOS spectrum over
$-4000<v<4000\,{\rm km\,s^{-1}}$, with the continuum solution held fixed.
The MODS $W_{\rm H\alpha}$ and GMOS emission-line profile were obtained at different epochs;
combining them assumes that broad-line variability does not materially alter
the normalization--shape comparison.

The spectral model consisted of a flat local continuum, the continuous
\textsc{Cloudy}-weighted stratified BLR of Eq.~(\ref{eq:blr_radial}), one
narrow H$\alpha$ emission component, the [N~{\sc ii}] emission doublet, and an effective
partial-covering absorber. We fitted $p$, $A_{\rm sub}$, $s_{\rm sub}$, and
the broad-line velocity offset together with the nuisance parameters
describing the narrow emission and absorption. The
parameters $p$ and $f_{\rm vir}$ are strongly covariant, because both alter
the relative prominence of the core and wings. To reduce this degeneracy, we
fixed the effective virial normalization to its natural value
$f_{\rm vir}=1$ rather than treating it as an additional free parameter, and
fitted $p$. The Pa$\alpha$ and H$\beta$ profiles were not included
in the likelihood; both are predictions of the H$\alpha$-constrained model.
The emission-line FWHMs quoted below were measured directly from these predicted
integrated profiles. Figure~\ref{fig:halpha_profile} shows the resulting
decomposition, whose best-fitting parameters are
%---------------------------------------------------------------------
\begin{equation}
 \begin{aligned}
 p&=1.55, & A_{\rm sub}&=1.59, & s_{\rm sub}&=0.39,\\
 f_{\rm vir}&\equiv1, &
 \Delta v_{\rm BLR}&=+17\,{\rm km\,s^{-1}}.&&
 \end{aligned}
 \label{eq:halpha_best}
\end{equation}
%---------------------------------------------------------------------
The fitted $p=1.55$ implies
$dC_{\rm BLR}/d\ln r\propto r^{-0.55}$ well inside $r_{\rm sub}$. The
pile-up contributes 7\% of the integrated covering but 22\% of the H$\alpha$
luminosity. Its emissivity-weighted radius is 1.9~pc and its predicted
${\rm FWHM}=552\,{\rm km\,s^{-1}}$. The complete continuous emission profile predicts
%---------------------------------------------------------------------
\begin{align}
 {\rm FWHM}_{\rm H\alpha,b}&=814\,{\rm km\,s^{-1}},\nonumber\\
 {\rm FWHM}_{\rm Pa\alpha,b}&=824\,{\rm km\,s^{-1}}; \qquad \chi_\nu^2=1.07.
 \label{eq:halpha_width}
\end{align}
%---------------------------------------------------------------------
The model therefore reproduces the observed exponential-like profile
without any electron-scattering contribution.
Importantly, the data require no additional Gaussian or physically
distinct intermediate-line region: the H$\alpha$ core and shoulders arise
within the continuous stratified BLR. Intermediate-width permitted-line
components have been reported in ordinary AGNs, but whether they form a
separate region or the outer part of a continuous BLR remains debated
\citep{Adhikari2017}. The predicted total Pa$\alpha$ FWHM is also close to the
published $819\pm24\,\mathrm{km\,s^{-1}}$ value \citep{Lin2026}, although
structured core residuals remain.

The fitted covering slope remains somewhat flatter than the
$p\simeq2.2\pm0.5$ range
favored by some reverberation-mapping applications of related global BLR
models \citep{Netzer2020}. This comparison is not one-to-one because our
radial weighting also includes the angle-dependent continuum and the
nonlinear \textsc{Cloudy} emissivity. More directly, the model predicts
$R_{50}({\rm H\alpha})/R_{50}({\rm H\beta})=1.6$, consistent with the
moderate Balmer-line stratification measured by reverberation mapping
\citep{Bentz2010,Feng2021}.

\subsection{Balmer and Paschen decrements}
\label{sec:balmer_decrement}

With the BLR structure specified by the H$\alpha$ analysis, the integrated
hydrogen-line decrements are predictions rather than inputs to the continuum
fit. The preferred density-transition model predicts an intrinsic
H$\alpha$/H$\beta$ ratio of 5.2. Attenuation by the same dusty cloud required
by the continuum gives
%---------------------------------------------------------------------
\begin{align}
 \left(\frac{F_{\rm H\alpha}}{F_{\rm H\beta}}\right)_{\rm int}
 &=5.2,\nonumber\\
 \left(\frac{F_{\rm H\alpha}}{F_{\rm H\beta}}\right)_{\rm att}
 &=5.2\times 10^{0.4(A_{\rm H\beta}-A_{\rm H\alpha})}=17.6.
 \label{eq:balmer_decrement_prediction}
\end{align}
%---------------------------------------------------------------------
The attenuated prediction agrees with the broad-line value $16.9\pm1.8$
measured by \citet{Lin2026}, using the same smooth CCM reddening curve as the
continuum fit. The large decrement follows from non-Case-B line formation in 
dense gas combined with dust extinction, rather than from either effect alone.

As a posterior diagnostic, the same model predicts
Pa$\alpha$/Pa$\gamma\simeq5.2$ after attenuation, compared with
$7.46\pm0.33$ \citep{Lin2026}; it therefore underpredicts the relative
Paschen decrement by about 30\%. Pa$\beta$ was not used because its published
broad flux is affected by strong telluric absorption. The absolute
Pa$\alpha$ normalization also remains too large for the baseline inner-BLR
contribution by a factor of order three.
Increasing the extinction could steepen Pa$\alpha$/Pa$\gamma$ toward the
observed value, but would also increase Pa$\alpha$/H$\alpha$ and worsen the
absolute Pa$\alpha$ excess. The two discrepancies therefore cannot be
simultaneously removed by a single rescaling of the covering factor or
foreground attenuation. High-redshift nebular attenuation curves can be
steeper than CCM in the near-infrared and vary substantially among objects
\citep{Sanders2025Attenuation,Reddy2026Attenuation}, adding systematic
uncertainty to intrinsic Paschen ratios without resolving both discrepancies.
\citet{Netzer2020} found that photoionization calculations can
miss the absolute Balmer and Paschen luminosities by factors of 2--5,
highlighting the continuing uncertainty in hydrogen-line transfer through
dense, optically thick BLR gas.

\FloatBarrier

%=====================================================================
\section{Balmer-series profiles and absorption}
\label{sec:balmer_profiles}

\subsection{Predicted H$\beta$--H$\delta$ profiles}
\label{sec:balmer_stratification}

\begin{figure}[!t]
\centering
\includegraphics[width=0.95\columnwidth]{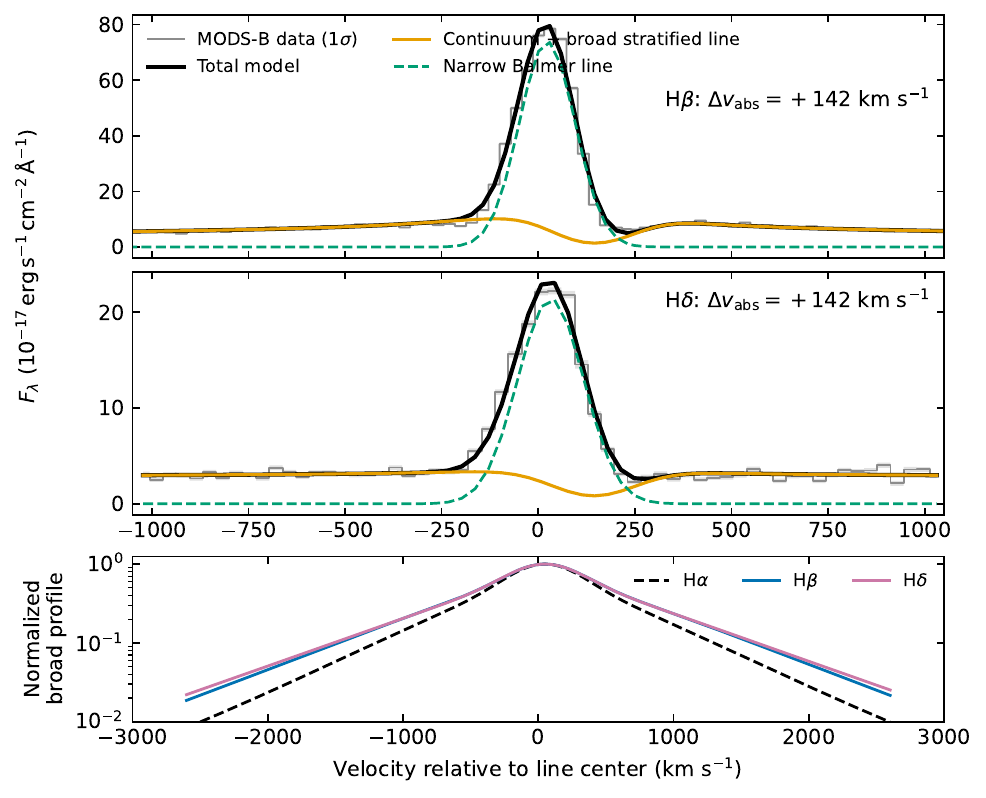}
\caption{Diagnostic fits to the LBT/MODS-B H$\beta$ and H$\delta$ spectra
\citep{Lin2026}. Gray shows the data, black the total model, orange the
absorbed continuum plus broad emission, and green the narrow emission. Broad
shapes are fixed by H$\alpha$ and normalizations by its flux and the
attenuated \textsc{Cloudy} ratios. Both panels share $C_{\rm los}=1$,
$\Delta v_{\rm abs}=+142\,{\rm km\,s^{-1}}$, and
$\sigma_\tau=22\,{\rm km\,s^{-1}}$; only local continua and narrow emission
were fitted. The lower strip compares the normalized unabsorbed broad
profiles.}
\label{fig:balmer-profiles}
\end{figure}

The H$\alpha$-profile analysis also fixes the intrinsic higher-order Balmer
profiles and their relative luminosities. Once the disk SED, cloud properties,
radial boundaries, density transition, and radial covering distribution have
been specified, the
\textsc{Cloudy}-weighted distributions in radius and ionization parameter
are predictions of the same model. Denoting the emissivity-weighted half-light
radius by $R_{50}$, the complete continuous distribution, marginalized over
the BLR illumination angles gives
%---------------------------------------------------------------------
\begin{align}
 R_{50}({\rm H\alpha})&=0.27~{\rm pc}, &
 \log U_{50}({\rm H\alpha})&=-1.2,\nonumber\\
 R_{50}({\rm H\beta})&=0.17~{\rm pc}, &
 \log U_{50}({\rm H\beta})&=-0.8,\nonumber\\
 R_{50}({\rm H\delta})&=0.17~{\rm pc}, &
 \log U_{50}({\rm H\delta})&=-0.8.
 \label{eq:balmer_radii}
\end{align}
%---------------------------------------------------------------------
Because $Q_{\rm H}$ is anisotropic, a given radius does not correspond to a
single value of $U$. Here, $\log U_{50}$ denotes the median of the complete
line-emissivity-weighted $\log U$ distribution over
$65^\circ\leq i_{\rm BLR}\leq85^\circ$, rather than the value at one
selected illumination angle. The higher-order lines cluster at
$0.17$~pc, while H$\alpha$ receives appreciably more emission from larger
radii.

These quantities describe the complete continuous distribution and are not constrained by the
narrow Balmer absorption discussed below. They do not depend on $f_{\rm vir}$,
which only converts radius into velocity. The predicted H$\alpha$-to-H$\beta$
radius ratio is close to the weighted reverberation-lag ratio
$\Delta t({\rm H\alpha})/\Delta t({\rm H\beta})=1.54$ found by the Lick AGN
Monitoring Project and modestly larger than the ratio 1.27 measured in
NGC~2617 \citep{Feng2021,Bentz2010}. H$\gamma$ has the same predicted
half-light radius as H$\delta$ at the quoted precision; their predicted
integrated FWHMs are correspondingly about 10\% larger than that of the stratified
H$\alpha$ component, although each width depends on the full
emissivity-weighted radial mixture rather than on $R_{50}$ alone.
The predicted $R_{50}({\rm H\beta})=0.17$~pc corresponds to a characteristic
light-travel time of approximately 200 rest-frame days, substantially longer
than the 45--50 days expected from the standard radius--luminosity relation
\citep{Bentz2013}. The difference may reflect that an emissivity-weighted
half-light radius is not a responsivity-weighted lag, as well as the adopted
radial covering profile and nonstandard super-Eddington ionizing SED.

With no further adjustment of the BLR parameters, the model predicts
%---------------------------------------------------------------------
\begin{align}
 {\rm FWHM}_{\rm H\beta,b}&=1430\,{\rm km\,s^{-1}},\nonumber\\
 {\rm FWHM}_{\rm H\delta,b}&=1410\,{\rm km\,s^{-1}}.
 \label{eq:balmer_width_predictions}
\end{align}
%---------------------------------------------------------------------
The H$\beta$ prediction agrees with the $1402\pm35\,{\rm km\,s^{-1}}$ reported by
\citet{Lin2026}. The reported width depends on the adopted multicomponent
decomposition. We therefore retain H$\beta$ as a prediction and diagnostic
rather than allowing it to determine $p$.

As a direct forward test, we fixed the H$\alpha$-constrained broad-emission
profile shapes and normalizations from H$\beta$ through H$\delta$ to the
\textsc{Cloudy} predictions, and tested the outer pile-up separately below.
We refitted only each local continuum, additive
unabsorbed narrow emission, and one effective absorber. Figure~\ref{fig:balmer-profiles}
shows that this model follows the overall profiles and broad wings without
independently rescaling any higher-order broad line. The formal reduced
chi-square values are 6.1, 7.4, and 3.5 for H$\beta$, H$\gamma$, and
H$\delta$, respectively. They are elevated by correlated residuals where
the unresolved narrow emission and absorption overlap, rather than by a
systematic failure in the broad wings.

The dense outer pile-up that reproduces the H$\alpha$ emission core and shoulders has
no detectable H$\beta$--H$\delta$ emission counterpart. The upper limit on
pile-up H$\beta$ emission implies an observed
H$\alpha$/H$\beta\gtrsim60$ for this component. Fitting a pile-up Pa$\alpha$
emission component at the H$\alpha$-predicted centroid and width likewise gives an observed
Pa$\alpha$/H$\beta\gtrsim17$.  These ratios may reflect line-selective transfer in 
dense, optically thick gas, as independently suggested by the H$\alpha$ refill in
Sect.~\ref{sec:missing_halpha}, although extra extinction local to the pile-up
would act in the same direction. Such local extinction is plausible near
the dust sublimation and formation boundary. The unmodified \textsc{Cloudy} ratios from
one uniform zone are insufficient.

\subsection{Effective Balmer absorption}
\label{sec:balmer_absorption}

Dense, high-column BLR-like gas naturally produces Balmer absorption when
radiative trapping and collisions populate the $n=2$ level of neutral
hydrogen \citep{Inayoshi2025,Yan2025}. This motivates locating the absorbers
within or adjacent to the stratified BLR, but does not determine their geometry.
Geometry may be at least as important as microphysics: \citet{Shangguan2026}
found Balmer absorption in only $\simeq0.05\%$ of ordinary type-1 AGNs, rising
to $\simeq10\%$ in the high-Eddington-ratio, weak-Fe\,II subpopulation.
An independent DESI selection found 15 Balmer-absorbed objects
among 54,556 broad-H$\alpha$ QSOs, a similar incidence of approximately
$0.03\%$ \citep{Ji2026b}. These population incidences measure how rarely a
sightline intercepts Balmer-absorbing gas, not the fraction of the source
covered within an intercepted object. The rarity may also reflect
selection: in ordinary dust-rich AGNs, sightlines through BLR clouds may be
strongly obscured and therefore absent from optically selected type-1 samples.

Lord/Egg is one such rare Balmer-absorbed broad-line AGN. Its common redshifted Balmer
absorber has a line-of-sight covering fraction consistent with unity. The
$i\simeq73\degr$ sightline may therefore graze a flattened BLR where absorbing
clumps collectively cover most of the projected Balmer-emitting region. In a
simple uniform circular geometry,
$R_{\rm abs}\gtrsim C_{\rm los}^{1/2}R_{\rm em}\simeq R_{\rm em}$,
corresponding to a transverse scale of order $0.1$--$0.2$~pc for the inferred
Balmer half-light radii. This transverse extent should not be identified with
the much smaller radial thickness $N_{\rm H}^{\rm c}/n_{\rm H}$ of an
individual \textsc{Cloudy} cloud: the effective absorber more plausibly
represents a flattened structure or the cumulative projected area of many smaller
clumps.

We characterized the Balmer absorption troughs independently of the broad-line-width
diagnostic using a phenomenological decomposition. We fixed the narrow-line
kinematics to the emission fits and instrumental line-spread function, while
allowing line-specific local continua and narrow-line amplitudes as nuisance
parameters. The spectra require a narrow blueshifted
H$\alpha$ trough and redshifted troughs beside the unresolved H$\beta$--H$\delta$
cores. For a partially covered source, the transmitted profile is
%---------------------------------------------------------------------
\begin{align}
 F_i(v)&=F_{{\rm n},i}(v)+F_{0,i}(v)
 \left\{1-C_{{\rm los},i}
 \left[1-e^{-\tau_i(v)}\right]\right\},\nonumber\\
 \tau_i(v)&=\tau_{0,i}
 \exp\left[-\frac{(v-v_i)^2}{2\sigma_{\tau,i}^2}\right].
 \label{eq:partial_covering}
\end{align}
%---------------------------------------------------------------------
Here $F_{0,i}$ is the reconstructed continuum plus broad-line emission and
$F_{{\rm n},i}$ is additive unabsorbed narrow emission. We take
$C_{{\rm los},i}$ to be velocity independent. For the common red absorber,
$C_{{\rm los},i}=C_{\rm los}$, $v_i=v_{\rm red}$,
$\sigma_{\tau,i}=\sigma_\tau$, and
$\tau_{0,i}/\tau_{0,{\rm H\beta}}=f_i\lambda_i/
(f_{\rm H\beta}\lambda_{\rm H\beta})$; the blueshifted H$\alpha$ absorber is
fitted separately. The representative decomposition gave

%---------------------------------------------------------------------
\begin{align}
 \Delta v_{\rm abs}({\rm H\alpha})&=-130~{\rm km\,s^{-1}}, &
 {\rm FWHM}_{{\rm abs,obs}}&=134~{\rm km\,s^{-1}},\nonumber\\
 \Delta v_{\rm abs}({\rm H\beta})&=+140~{\rm km\,s^{-1}}, &
 {\rm FWHM}_{{\rm abs,obs}}&=170~{\rm km\,s^{-1}},\nonumber\\
 \Delta v_{\rm abs}({\rm H\delta})&=+170~{\rm km\,s^{-1}}, &
 {\rm FWHM}_{{\rm abs,obs}}&=200~{\rm km\,s^{-1}}.
 \label{eq:balmer_absorbers}
\end{align}
%---------------------------------------------------------------------
After convolution with the instrumental line-spread function, the common-red
fit gives
$C_{\rm los}\simeq1$, $v_{\rm red}=+142~{\rm km\,s^{-1}}$, and
$\sigma_\tau\simeq22~{\rm km\,s^{-1}}$; saturation leaves only a lower limit
on the line-center optical depth. The red H$\beta$--H$\delta$ troughs and
narrow cores are unresolved at the wavelength-dependent MODS-B resolution
calibrated from [O~{\sc iii}]~$\lambda5008$; their observed FWHMs of
$170$--$200$ and $170$--$217~{\rm km\,s^{-1}}$, respectively, are
instrumental, with narrow H$\delta$ at most marginally resolved. The absorber
centroids agree within $30~{\rm km\,s^{-1}}$, well below one resolution
element, and provide no
resolved evidence for a systematic velocity shift with Balmer order.

Quoting absorption equivalent widths as positive magnitudes, we obtain
$W_{\rm abs}({\rm H\alpha})=2.3$~\AA\ and
$W_{\rm abs}({\rm H\beta})=2.8$~\AA, consistent in scale with the published
decomposition. A simultaneous H9--H12 curve-of-growth fit gives
$N({\rm H},n=2)\simeq(2$--$5)\times10^{15}\,\mathrm{cm^{-2}}$
(Appendix~\ref{app:high_order_balmer}). H8 was excluded because it is blended
with He~{\sc i}~$\lambda3889$. Correlated residuals remain in the unresolved
core structure.
H$\alpha$ is P~Cygni-like, but its $120$--$130\,\mathrm{km\,s^{-1}}$
blueshift implies modest outward motion rather than a fast outflow. 
For this blueshifted absorber, the corresponding H$\beta$ optical depth is smaller by the atomic
factor 7.25 and any counterpart would overlap the unresolved H$\beta$ core,
so its nondetection is not constraining. The blueshifted H$\alpha$ absorber
covers the continuum source and the stratified broad-line emission; narrow
H$\alpha$ and [N~{\sc ii}] emission are additive, unabsorbed terms, so
the spatially extended narrow-line region (NLR) is largely uncovered.
We tentatively associate the blueshifted H$\alpha$ absorber with the outer-BLR
complex near 2~pc, possibly with the warm CaT-bearing substructure modeled in
Sect.~\ref{sec:cat_absorption}. The H$\alpha$ and CaT centroids agree within
one resolution element, although their different preferred covering fractions
imply distinct projected substructure. In this geometry, H$\alpha$ emission
integrates the complex's full velocity field, whereas absorption selects its
narrow foreground projection; this is a working physical identification
rather than a unique inference from the profiles. H$\beta$--H$\delta$ are instead
inverse-P~Cygni-like. Their
consistent $+140$--$+170\,\mathrm{km\,s^{-1}}$ centroids favor a common
redshifted absorber and could indicate modest infall, but their intrinsic
widths remain unresolved and the inferred depths depend on decomposition of
the narrow cores. A higher-velocity contrast is the variable Balmer-absorbed
QSO J1157, where \citet{Ji2026b} inferred BLR-scale inflow at several hundred
to approximately $1000\,\mathrm{km\,s^{-1}}$ and modeled possible
Ca~{\sc ii} H and K variability with common Balmer kinematics. 
Lord/Egg, which is also less massive and less luminous, lacks such fast
absorption; the
comparison suggests, but does not establish, a shared low-ionization/Balmer
phase.

\subsection{\texorpdfstring{The missing red H$\alpha$
absorption}{The missing red Halpha absorption}}
\label{sec:missing_halpha}

The absence of the expected redshifted H$\alpha$ counterpart to the
H$\beta$--H$\delta$ troughs is not unique to Lord/Egg. Departures from the
atomic Balmer optical-depth ratios also occur in Balmer-absorbed broad
absorption line quasars, where nonuniform or transition-dependent covering is
usually invoked \citep{Leighly2025Balmer}. \citet{Hall2007} noted that emission
by the absorber itself can refill H$\alpha$ more strongly than H$\beta$, and
\citet{Liu2026Balmer} recently formulated the analogous ``Balmer absorber
strength problem'' for LRDs in terms of the line source function. The velocity
sign is not universal: classical Balmer absorption troughs are generally
blueshifted, whereas \citet{Zhou2019Inflow} identified redshifted Balmer and
metastable-helium absorbers at BLR scales. The issue addressed here is therefore
the transition-dependent trough strength, not the redshift of the gas.

One trough constrains $C_{\rm los}(1-e^{-\tau})$, not $C_{\rm los}$ and
$\tau$ separately. For the same $n=2$ column and velocity width, atomic line
strengths give $\tau_0({\rm H\alpha})/\tau_0({\rm H\beta})=7.25$. The
redshifted absorber detected from H$\beta$ through H$\delta$ should therefore
produce still stronger H$\alpha$ absorption if it covers the Balmer emission
regions similarly. Instead, no H$\alpha$ trough is detected at its
$+140$--$170\,\mathrm{km\,s^{-1}}$ velocity.
We tested whether radial stratification could create the required differential
covering. At $+146\,\mathrm{km\,s^{-1}}$, the complete preferred model,
including the outer pile-up, gives $R_{50}({\rm H\alpha})=0.80$~pc and
$R_{50}({\rm H\beta})=0.52$~pc, or a ratio of 1.54, close to the integrated
value of 1.6. Because the line emissivities are locally close to power laws,
selecting a projected velocity shifts their characteristic radii nearly
homologously; the pile-up only modestly breaks this scaling. A centered screen
that covers 0.80 of the H$\beta$ surface brightness covers 0.74 of H$\alpha$
and approximately 0.81 of H$\gamma$ and H$\delta$. Thus the continuous
stratified BLR does not yield
$C_{{\rm los},{\rm H\alpha}}\ll C_{{\rm los},{\rm H\beta}}$ at the
absorber velocity.

We therefore allowed the absorbing gas to return a fraction of the line flux
removed in each transition to the line of sight.
If $F_{0,i}(v)$ denotes the unabsorbed continuum plus broad-line emission profile behind the
absorber, the flux removed in transition $i$ is
%---------------------------------------------------------------------
\begin{equation}
 {\cal A}_i(v)=F_{0,i}(v)C_{\rm los}
 \left[1-e^{-\tau_i(v)}\right].
 \label{eq:removed_flux}
\end{equation}
%---------------------------------------------------------------------
We write the observed profile as
%---------------------------------------------------------------------
\begin{equation}
 F_i(v)=F_{{\rm n},i}(v)+F_{0,i}(v)-{\cal A}_i(v)
       +\eta_i\left[{\cal A}_i\ast G(\sigma_r)\right](v),
 \label{eq:balmer_refill}
\end{equation}
%---------------------------------------------------------------------
where $F_{{\rm n},i}$ is uncovered narrow emission, $\ast$ denotes velocity
convolution, and $G$ is a normalized Gaussian redistribution kernel of width
$\sigma_r$. Thus $\eta_i=0$ gives
pure absorption, while $\eta_i=1$ and $\sigma_r\rightarrow0$ return exactly
the flux removed in each velocity channel. Values above unity would indicate
net local line emission in addition to scattering. This phenomenological
returned-flux prescription is motivated by absorber self-emission and
source-function treatments \citep{Hall2007,Liu2026Balmer}, but does not assume
which process supplies the returned photons. In optically thick gas,
trapped Ly$\beta$ photons and cascades can feed H$\alpha$, whereas
absorbed H$\beta$ photons can be redistributed through Pa$\alpha$; H$\alpha$
can consequently have a much larger effective source function. 

With one velocity-independent covering fraction shared by H$\alpha$--H$\delta$,
the diagnostic fit gives $(C_{\rm los},v_{\rm red},\sigma_\tau)=(1.0,
+142~{\rm km\,s^{-1}},21.9~{\rm km\,s^{-1}})$, $\eta_{\rm H\alpha}=0.98$, 
no higher-order refill, and $\sigma_r\lesssim22\,\mathrm{km\,s^{-1}}$. 
The returned component is therefore
coherent in velocity to within the instrumental resolution. As a control, the
separate blueshifted H$\alpha$ absorber gives only $\eta=0.03$. The joint
fit is limited by unresolved core residuals, so these
values establish the required line dependence rather than precise confidence
intervals. If the H$\alpha$ covering is instead left completely free, a
no-refill fit of comparable quality drives it to nearly zero. The data alone
therefore permit an extreme differential-covering solution. Under the covering
relation predicted by the preferred stratified BLR, however, our model requires
velocity-coherent H$\alpha$ refill.
This narrow returned component may be associated with the H$\alpha$ excess
from the outer pile-up; its narrower profile reflects the restricted foreground
projection, whereas the observed pile-up emission integrates the full complex.

The radial BLR grid used elsewhere in this paper was calculated with the C23
release of \textsc{Cloudy}. Although C23 includes diffuse emission, that grid
was constructed to predict BLR emissivities and was not configured to return
the absorber's diffuse and scattered line intensity into the observed ray. To
test the mechanism proposed by \citet{Liu2026Balmer}, we therefore performed a
separate diagnostic calculation with the C25 release of \textsc{Cloudy}
\citep{Gunasekera2025}, supplemented by their reciprocal continuum-shielding
prescription. We followed the spectrum from the observer-facing, shielded side
of a representative one-dimensional BLR cloud. The calculation gives
returned-to-removed intensity ratios of order 0.9 for H$\alpha$ and 0.1 for
H$\beta$, qualitatively matching the required line dependence. It is used as
a physical consistency check rather than as a prior on $\eta_i$: a
self-consistent calculation must include the local continuum plus broad-line
radiation field and three-dimensional angular transfer.

\citet{Wang2026Balmer} found redward centroid shifts from H$\alpha$ to
higher Balmer transitions in two LRDs, while Ca~{\sc ii} K followed
H$\alpha$, indicating that optically thick, velocity-stratified gas can make
different transitions sample different depths. Lord/Egg lacks a resolved
monotonic shift within H$\beta$--H$\delta$, but the coexistence of blueshifted
H$\alpha$/CaT and redshifted higher-Balmer absorption is qualitatively
consistent with the optically thick, velocity-stratified failed-wind complex
developed in Sect.~\ref{sec:absorber_kinematics}. Our transfer calculation does
not yet predict the transition-dependent centroids.

%=====================================================================
\section{Weak broad high-excitation emission}
\label{sec:high_excitation}

The same angle-dependent disk SED and stratified \textsc{Cloudy} calculation
predict only weak broad high-excitation emission. In particular, the
preferred model gives
%---------------------------------------------------------------------
\begin{equation}
 W_{\rm b}(\mathrm{He\,II}\,\lambda4686)\simeq
 2.3\,\text{\AA}.
 \label{eq:heii_ew}
\end{equation}
%---------------------------------------------------------------------
\citet{Lin2026} did not detect a secure broad He~{\sc ii} component. The weak
broad component predicted by our model is consistent with this nondetection.
The remaining broad high-excitation lines accessible in the optical are
predicted to have EWs of order $1\,\text{\AA}$ or less. Neither
\citet{Ji2025z01} nor \citet{Lin2026} reported line-by-line EW
limits for these nondetections, so we did not include them in the fit; they
provide an a posteriori check on the hard ionizing continuum seen by the BLR.

%=====================================================================
\section{Low-ionization helium and metal absorption}
\label{sec:metal_absorption}

We interpret the absorbers as a multiphase continuation of the stratified BLR,
not unrelated screens. Ionized BLR surfaces supply He~{\sc i}$^*$ absorption;
near the outer boundary, the pile-up contributes H$\alpha$ emission and its
foreground projections produce Balmer absorption, locally heated dust-poor
interfaces produce CaT absorption, and cool dusty cores produce Na~{\sc i}
and K~{\sc i} absorption. These are zones within distributions
of radius, density, column, dust content, and covering rather than discrete
shells.

\subsection{Metastable He~{\sc i} absorption}
\label{sec:hei_absorption}

All three local LRDs of \citet{Lin2026} show blueshifted
He~{\sc i}~$\lambda10833$ absorption. For Lord/Egg, a three-component FIRE fit
gives velocities $(-514,-440,-154)\,\mathrm{km\,s^{-1}}$, $\tau_0\gg1$, and
projected coverings $(0.41,0.67,0.67)$, so the saturated trough mainly
constrains covering. The broad profile has
$\mathrm{FWHM}=1080\,\mathrm{km\,s^{-1}}$ and fitted $W_{\rm em}=57$~\AA,
versus the \textsc{Cloudy} prediction of 210~\AA; its normalization was
therefore free. Ionized surface layers with
$\log(N_{\rm H}/\mathrm{cm^{-2}})=21.3$--21.5 already saturate
He~{\sc i}$^*$ while remaining Balmer transparent, allowing the two fastest
components to trace radiation-accelerated BLR skins rather than separate
screens or a dense wind (Appendix~\ref{app:hei_absorption}).

\subsection{A common Na~{\sc i} and K~{\sc i} absorber}
\label{sec:alkali_absorber}

\begin{figure}[!t]
 \centering
 \includegraphics[width=0.99\columnwidth]{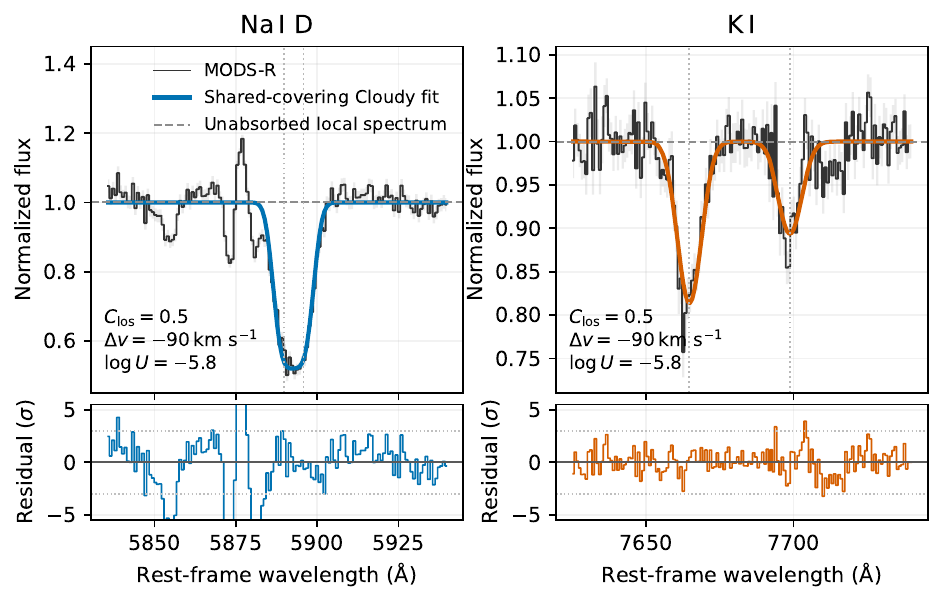}
 \caption{Continuum-normalized MODS-R Na~{\sc i}~D and K~{\sc i} profiles
 \citep{Lin2026} and the shared-covering dusty \textsc{Cloudy} model. Its
 ionization parameter, common $C_{\rm los}$, and centroid were fitted jointly;
 doublet ratios and intrinsic FWHMs were fixed before convolution to the
 instrumental resolution. Dotted lines mark the fitted centroids.}
 \label{fig:alkali_profiles}
\end{figure}

Lord/Egg shows exceptionally strong Na~{\sc i}~D, K~{\sc i}, and
Ca~{\sc ii}-triplet absorption. \citet{Lin2026} measured the summed
rest-frame absorption equivalent widths (EWs) and intrinsic absorption-profile FWHMs
%---------------------------------------------------------------------
\begin{align}
 (W_{\rm NaD},W_{\rm K},W_{\rm CaT})
   &=(6.6\pm0.4,3.1\pm0.1,16.7\pm0.5)\,\text{\AA},\nonumber\\
 \mathrm{FWHM}_{\mathrm{Na,K,CaT}}
   &=(337,296,441)\,\mathrm{km\,s^{-1}},
 \label{eq:metal_abs_observed}
\end{align}
%---------------------------------------------------------------------
with centroid offsets of $-98$, $-101$, and $-61\,\mathrm{km\,s^{-1}}$,
respectively. The nearly identical Na~{\sc i} and K~{\sc i} centroids and
comparable widths favor one cool absorber, modeled as a dusty zone of the
stratified BLR with common \textsc{Cloudy} ion columns, projected covering,
and line-of-sight centroid.
We retained the measured widths because optical depth and unresolved velocity
substructure can weight the doublets differently. As for Balmer absorption,
these restricted foreground projections need not sample the virial emission
kernel, so their widths do not determine radii. Their common, modest
blueshifts favor a weak bulk outflow, while random motions may dominate the
line widths.

Their placement at a few $r_{\rm sub}$ instead follows from ionization and
dust survival. The Na~{\sc i} and K~{\sc i} thresholds, 5.14 and 4.34~eV,
lie below the Lyman edge, so neutral hydrogen cannot shield them from the
optical and near-ultraviolet continuum. Dense clouds beyond $r_{\rm sub}$
provide low $U$ and internal dust filtering, producing cool neutral interiors.
We explored $-6.5\leq\log U\leq-5.5$, or
$2.8\lesssim r/r_{\rm sub}\lesssim8.8$ at the adopted density. Rather than
integrating a full radial cloud distribution with position-dependent velocity
and covering, we represented both doublets by one characteristic complex.
Its radius and covering fraction are therefore effective properties of the
neutral substructure, not the mean radius or sky covering of all dust.

We tested this picture using constant-density \textsc{Cloudy} models of cool,
dusty clouds exposed to the disk SED at $i_{\rm BLR}=73\degr$, each with
$N_{\rm H}^{\rm c}=10^{23}\,\mathrm{cm^{-2}}$ and
$Z=0.1\,Z_\odot$. The adopted ISM dust-to-gas ratio was 0.05 of the
Galactic value, approximately half a linear metallicity scaling; for this
column it gives $A_V^{\rm c}\simeq2.8$~mag, close to the continuum result.
The standard \textsc{Cloudy} interstellar-medium depletion
prescription left 20\% of Na and 30\% of K in the gas phase ($-0.7$ and
$-0.5$ dex; \citealt{ChaffeeWhite1982,Chatzikos2023}). The preferred clump,
with $n_{\rm H}\sim10^{11}\,\mathrm{cm^{-3}}$ and $\log U\simeq-5.8$,
lies at $r\simeq3.9\,r_{\rm sub}$. It belongs to the dense outer-cloud
population introduced in Sect.~\ref{sec:halpha}; its temperature falls from
approximately $10^4$ to $300$~K through the cloud. Dust filters the directly
incident sub-Lyman continuum through each cloud. The shielded interior gives
$N(\mathrm{Na\,I})=9.9\times10^{14}\,\mathrm{cm^{-2}}$ and
$N(\mathrm{K\,I})=1.4\times10^{13}\,\mathrm{cm^{-2}}$.

We fitted the MODS-R absorption profiles jointly, sharing $n_{\rm H}$,
$N_{\rm H}^{\rm c}$, dust content, metal depletion, $U$, covering fraction,
and centroid.
The Cloudy columns, atomic optical-depth ratios, and measured intrinsic FWHMs
were fixed at each grid point; the transmitted spectra were convolved to the
MODS-R resolution, while smooth local unabsorbed spectra were fitted. The
resulting absorption solution is
%---------------------------------------------------------------------
\begin{align}
 \log U&=-5.8,\qquad r=3.9\,r_{\rm sub},\qquad C_{\rm los}=0.5,\nonumber\\
 \Delta v&=-90~\mathrm{km\,s^{-1}},\qquad
 (W_{\rm NaD},W_{\rm K})
   =(6.0,3.0)~\text{\AA}.
 \label{eq:alkali_profile_fit}
\end{align}
%---------------------------------------------------------------------
The continuum fit attenuates the compact accretion-disk sightline by the full
cloud dust column. The alkali lines arise only in neutral substructures that
cover part of the projected optical source; thus $C_{\rm los}=0.5$ refers to
their projected line covering, not to the dust column or the cloud's global
sky covering.

Figure~\ref{fig:alkali_profiles}
shows that K~{\sc i} is reproduced particularly well and the principal
Na~{\sc i} trough is recovered; correlated residuals reflect neighboring weak
absorption and the one-component velocity field. The two ions can therefore
arise in dense, self-shielded material within the clumpy dusty component
beyond the main BLR, and their measured EWs refer to this cool phase alone.
Because both ions are resonance doublets, illuminated cool gas could scatter
continuum photons into the sightline. A wide-angle cool outflow would partly
fill the troughs near systemic velocity and possibly on their red sides
\citep{Prochaska2011,RupkeVeilleux2015}, but no such emission is evident.
This is compatible with a small global covering of the alkali-bearing
phase. Independently, the infrared energy budget gives $f_{\rm hd}\simeq0.02$,
although the total dust covering $C_{\rm dust}$ can be larger, despite
$C_{\rm los}\simeq0.5$ along our particular sightline.

\subsection{The Ca~{\sc ii} triplet}
\label{sec:cat_absorption}

In a sample of 17 JWST/NIRSpec grating spectra at $2<z<5$,
\citet{Greene2026CaT} detected CaT absorption in three objects and measured a
stacked $W_{\rm CaT}\simeq5$~\AA, attributing it to dense, warm gas associated
with the central engine. Lord/Egg stands out through the strength and
simultaneous detection of CaT, Na~{\sc i}, and K~{\sc i}, suggesting either a
favorable sightline through a multiphase complex or unusually strong
low-ionization structure near the BLR--torus transition. Among the three local
LRDs of \citet{Lin2026}, clear CaT absorption was detected only in Lord/Egg,
with $W_{\rm CaT}=16.7\pm0.5$~\AA, whereas the $z=5.3$ LRD Bz5.3 shows broad
CaT emission \citep{Tripodi2025a}. This contrast is consistent with geometric
stratification: emission integrates the BLR velocity field, while absorption
samples a restricted foreground projection.

Cool dusty clumps cannot produce the absorption: standard depletion gives
$W_{\rm CaT}\sim10^{-4}$~\AA, returning all calcium to the gas gives
$W_{\rm CaT}<0.6$~\AA, and its $\sim300$~K interior cannot populate the
metastable $3d\,^2D$ lower levels. Strong CaT instead requires dust-poor gas
at a few $10^3$~K with $n_e\gg10^7\,\mathrm{cm^{-3}}$
\citep{Liu2026Atmospheres}, resembling the standard AGN solution for strong
CaT emission from warm, deeply shielded, high-column neutral gas
\citep{FerlandPersson1989,Matsuoka2007,Panda2021}. Because Lord/Egg's soft,
high-inclination incident spectrum does not keep such gas warm radiatively, we
added uniform non-radiative heating to a grain-free, undepleted outer-cloud
model with the density and column adopted above and $\log U=-4.7$; H$^-$ and
free--free absorption were followed self-consistently.
Eq.~(\ref{eq:u_radius}) places this layer at
$1.1\,r_{\rm sub}\simeq1.9$~pc, where dust survival is marginal; shocks could
also erode grains, although we did not model that coupling.

Matching CaT fixes $C_{{\rm los},{\rm warm}}=0.35$ and gives
$W_{\rm CaT}=16.7$~\AA. The layer also contains saturated Na~{\sc i}, but its
overlap with the already saturated cool-phase absorption increases the
Na~{\sc i} column rather than the Na~{\sc i}~D EW; its weak K~{\sc i}
contribution is only $0.21$~\AA. This solution requires
$H_{\rm nr}=1.2\times10^{-5}\,\mathrm{erg\,cm^{-3}\,s^{-1}}$, only $0.05\%$
of the radiative heating at the illuminated face. Deeper into the cloud it
supplies approximately $85\%$ of the integrated heating and maintains
$T\simeq(7.7$--$8.0)\times10^3$~K. Heating diagnostics, energetics, and
abundance sensitivity are given in Appendix~\ref{app:cat_heating}.
Its Balmer opacity provides a physical basis for the possible association
with blueshifted H$\alpha$ absorption anticipated in
Sect.~\ref{sec:balmer_absorption}; the detailed H9--H12 test is given in
Appendix~\ref{app:high_order_balmer}. Additional Ca~{\sc ii} H and K and
Mg~{\sc i} absorption predictions for the warm layer are given in
Appendix~\ref{app:mg_absorption}.

We note that the optically thick atmosphere grid of
\citet{Liu2026Atmospheres} likewise cannot reproduce CaT and Na~{\sc i}~D
simultaneously. Its Lord/Egg model gives
$W_{\rm CaT}=16.9\pm0.3$~\AA\ and $W_{\rm NaD}=0.3\pm0.1$~\AA, and
$W_{\rm CaT}/W_{\rm NaD}\geq5$ throughout the grid, compared with the observed
ratio of 2.5. That model would therefore also require a multiphase
modification.

%=====================================================================
\section{Polarization}
\label{sec:polarization}

The VLT/FORS2 spectropolarimetry provides an independent geometric test of
the continuum solution. The continuum Stokes vector combines intrinsic
polarization from the electron-scattering disk atmosphere with dichroic
transmission through the foreground dust, whereas broad H$\alpha$ forms in
the BLR and receives only the dichroic contribution.
At the fixed SED parameters, the model gives
$p_{\rm cont}=1.62\%$, broad-H$\alpha$ polarization of $0.80\%$, and a
line-to-continuum position-angle offset $\Delta\vartheta=-52\degr$, close to
the observed $1.53\%$, $0.58$--$0.84\%$, and $-48\degr$. The
narrow H$\alpha$ and [N~{\sc ii}] cores are treated as unpolarized and excluded
from the likelihood; no additional polarized core component is fitted.
Appendix~\ref{app:polarization} gives the fixed inputs, Stokes model, and fit;
see also \citet{MadauMaiolino_POL_2026} for further details.

%=====================================================================
\section{\texorpdfstring{Absorber kinematics: outflow,
inflow, and a failed BLR wind}{Absorber kinematics: outflow, inflow, and a
failed BLR wind}}
\label{sec:absorber_kinematics}

\subsection{A velocity ladder}

Appendix~\ref{app:scorecard} (Table~\ref{tab:velocity_ladder}) compares
absorber centroids with escape speeds
at radii set by the emission or photoionization models, not by absorption
widths. The He~{\sc i}$^*$ entry uses its emissivity-weighted BLR radius, the
alkali and CaT radii follow from their ionization parameters, and the Balmer
absorbers are assigned the characteristic outer-pile-up radius. Absorption
selects foreground portions of the flattened BLR. At
$i=73\degr$, $v_{\rm LOS}\simeq0.96\,v_R+0.29\,v_z$; near-side outflowing gas
is therefore blueshifted, although random motions allow either sign.
Rotation across the extended BLR contributes only
$\sim25\,\mathrm{km\,s^{-1}}$ at 1.9~pc, so the assignments remain
suggestive rather than unique.

\begin{figure*}[!t]
\centering
\includegraphics[width=0.53\textwidth]{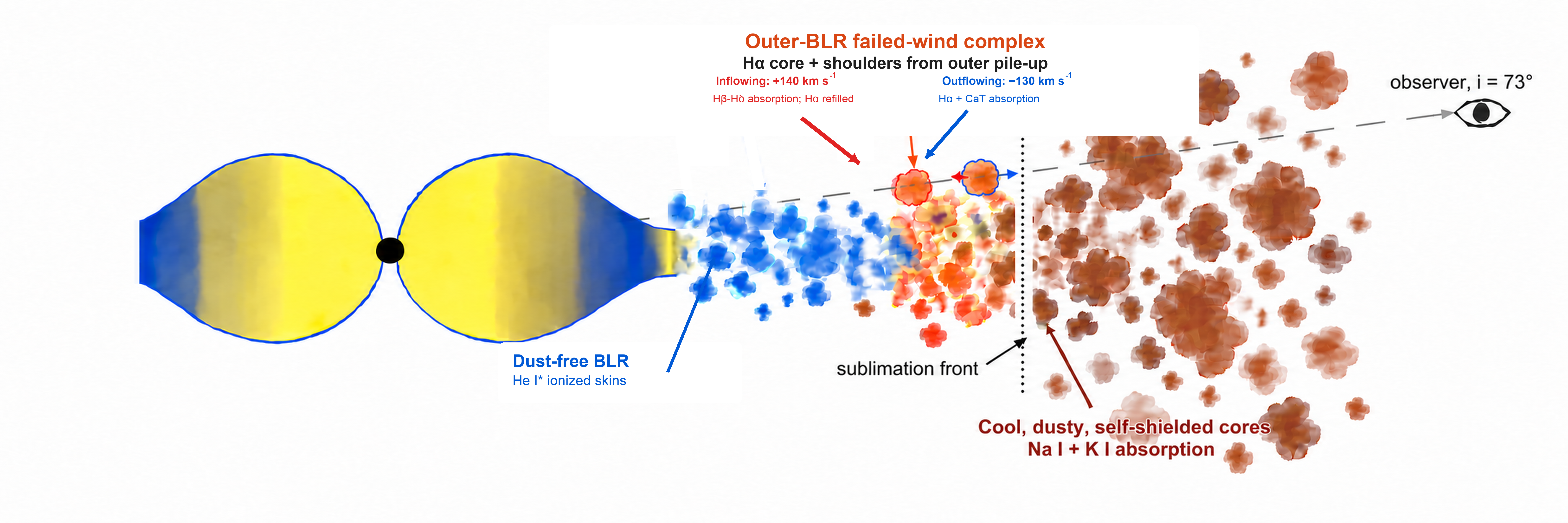}
\caption{Working geometry; sizes are not to scale. The orange outer-BLR
pile-up produces the H$\alpha$ core and shoulders. Outflowing foreground gas
produces H$\alpha$/CaT absorption; inflowing gas produces red H$\beta$--H$\delta$
absorption with H$\alpha$ refill. Inner ionized skins produce He~{\sc i}$^*$
absorption and distant dusty cores the alkalis. The dotted line marks dust
survival; the dashed line is the $i=73\degr$ sightline.}
\label{fig:blr_torus_transition}
\end{figure*}

\subsection{Bound outflow}

The low-column He~{\sc i} skins reach $-440$ and
$-514\,\mathrm{km\,s^{-1}}$, below the local
$910\,\mathrm{km\,s^{-1}}$ escape speed. H$\alpha$, CaT, and the alkalis trace
cooler gas at $0.2$--$0.5\,v_{\rm esc}$: H$\alpha$ and CaT are consistent with
bound outflow, while random motions may dominate the alkali widths.

This declining blueshift from ionized He~{\sc i}$^*$ skins to lower-ionization
phases is not peculiar to LRDs. In ordinary AGNs, higher-ionization broad
emission lines are more blueshifted than lower-ionization lines
\citep{Gaskell1982,Laor1997,Shen2016,Meyer2019}, while broad absorption line
(BAL) outflows likewise show ion-dependent trough kinematics
\citep{Gibson2009,Hamann2019BAL}.
The ionization--velocity sequence is consistent with a radially stratified,
decelerating outflow.

\subsection{Red Balmer absorption and infall}

The common H$\beta$--H$\delta$ absorber instead lies at
$+140$--$+170\,\mathrm{km\,s^{-1}}$ relative to systemic and
$+110$--$+145\,\mathrm{km\,s^{-1}}$ relative to the MODS narrow lines; the
instrumental reference difference cannot remove the offset. At 1.9~pc this
is projected infall at $0.4$--$0.5\,v_{\rm esc}$. The lack of red
He~{\sc i}$^*$ is consistent with softer illumination or different column
and covering, but does not distinguish among them.

Under the BLR covering relation, the missing red H$\alpha$ trough requires
velocity-coherent refill (Sect.~\ref{sec:missing_halpha}). We tentatively
associate the inflowing foreground sheet with the outer complex that produces
the nearly systemic H$\alpha$ pile-up: its integrated emission cannot cancel
a narrow trough, but local sheet emission can. A distinct Pa$\alpha$ feature
at $+140\,\mathrm{km\,s^{-1}}$ remains a prediction.

\subsection{A failed wind at the BLR--torus transition}

Coexisting outflowing and inflowing low-ionization gas is characteristic of failed
radiatively accelerated dusty outflows: exposed dust-driven material loses
its grains and falls back \citep{CzernyHryniewicz2011}. Multidimensional and
super-Eddington calculations similarly produce inner outflow and outer failed
trajectories \citep{NaddafCzerny2022,NaddafCzernyZajacek2022}. No calculation
has used Lord/Egg's $\dot m=80$ anisotropic SED; thick-disk shadowing makes
grain survival depend on $L(\theta)$ and may juxtapose dust-poor and dusty
gas. Figure~\ref{fig:blr_torus_transition} illustrates this interpretation,
with non-radiative heating confined to CaT-absorbing interfaces.

The absence of strong red He~{\sc i}$^*$ absorption and the predicted weak Pa$\alpha$
feature near $+140\,\mathrm{km\,s^{-1}}$ test this picture. A 0.2~pc source
crossed at a few $10^2\,\mathrm{km\,s^{-1}}$ evolves on $\sim10^3$~yr,
consistent with decade-long stability \citep{Burke2026Quiet}.

\subsection{Microphysics and limitations}
\label{sec:discussion}

The connected BLR--torus complex may be heated by dissipative collisions and
turbulence, as in clumpy-torus models
\citep{KrolikBegelman1988,Vollmer2004,Vollmer2022}, especially where
photoheating is attenuated. At $i=73\degr$, $H/R\gtrsim0.3$ can intercept the
nucleus while remaining compatible with $f_{\rm hd}=0.02$; colder emission
can raise the total dust covering.

The CaT calculation does not couple grain destruction and dissipation to
photoionization, line transfer, and covering; 8000~K gas alone does not prove
grain destruction. An analogous Mrk~231 model required compression from
He~{\sc i}$^*$-bearing H~{\sc ii} into dusty, partly neutral
Na~{\sc i}/Ca~{\sc ii} gas with reduced calcium depletion
\citep{Leighly2014}, although it modeled Ca~{\sc ii} H and K rather than CaT.

%=====================================================================
\section{Conclusions}
\label{sec:conclusions}

One connected, high-inclination super-Eddington geometry explains much of
Lord/Egg (Appendix~\ref{app:scorecard}). An attenuated disk plus young stars
fits the continuum; self-occultation explains the X-ray weakness; and
polarization supports the geometry. The same SED and stratified BLR reproduce
H$\alpha$, the Balmer decrement, and broad H$\beta$--H$\delta$, while supplying
metastable-He~{\sc i} opacity.

Low-ionization absorption traces the BLR--torus transition: heated dust-poor
interfaces produce CaT, the outer pile-up supplies H$\alpha$ and Balmer
absorption, and cool dusty layers dominate the alkalis. Lord/Egg is
exceptional because our sightline intercepts both phases. Such stratification
and failed dusty flows have AGN precedents
\citep{FerlandPersson1989,Leighly2014,CzernyHryniewicz2011}; all absorber
velocities remain below escape speed. Velocity-coherent re-emission preferably
explains the missing red H$\alpha$ trough, with higher Balmer photons converted
through Paschen channels; C25 supports this selectivity, although extreme
differential covering remains possible.

Open issues include the Pa$\alpha$/Pa$\gamma$ mismatch, the larger absolute
Pa$\alpha$ excess, and coupling CaT heating to grain destruction. Dense
electron-scattering envelopes may broaden inner lines \citep{Chon2026}, but
do not explain broader H$\beta$ than H$\alpha$, the Balmer decrement, or the
alkalis. Higher-resolution hydrogen spectra, far-infrared data, and
three-dimensional cloud transfer can test this picture.

\bibliographystyle{aa}
\bibliography{paper}

@ARTICLE{Adhikari2016,
       author = {{Adhikari}, T.~P. and {R{\'o}{\.z}a{\'n}ska}, A. and {Czerny}, B. and {Hryniewicz}, K. and {Ferland}, G.~J.},
        title = "{The Intermediate-Line Region in Active Galactic Nuclei}",
      journal = {\apj},
         year = 2016,
        month = nov,
       volume = {831},
          eid = {68},
        pages = {68},
          doi = {10.3847/0004-637X/831/1/68}
}

@ARTICLE{Adhikari2017,
       author = {{Adhikari}, T.~P. and {R{\'o}{\.z}a{\'n}ska}, A. and {Hryniewicz}, K. and {Czerny}, B. and {Ferland}, G.~J.},
        title = "{On the Intermediate Line Region in AGNs}",
      journal = {Frontiers in Astronomy and Space Sciences},
         year = 2017,
        month = sep,
       volume = {4},
          eid = {19},
        pages = {19},
          doi = {10.3389/fspas.2017.00019},
       adsurl = {https://ui.adsabs.harvard.edu/abs/2017FrASS...4...19A}
}

@ARTICLE{BaskinLaor2018,
       author = {{Baskin}, A. and {Laor}, A.},
        title = "{Dust inflated accretion disc as the origin of the broad line region in active galactic nuclei}",
      journal = {\mnras},
         year = 2018,
        month = feb,
       volume = {474},
        pages = {1970--1994},
          doi = {10.1093/mnras/stx2850}
}

@ARTICLE{Scholtz2026,
       author = {{Scholtz}, J. and {D'Eugenio}, F. and {Maiolino}, R. and {Brazzini}, M. and {{\"U}bler}, H. and {Ji}, X. and {Perna}, M. and {Sun}, F. and {Brocchi}, G. and {Carniani}, S. and {Cresci}, G. and {Ivey}, L.~R. and {Juod{\v{z}}balis}, I. and {Marconi}, A. and {Mazzolari}, G. and {Risaliti}, G. and {Trefoloni}, B.},
        title = "{Little Red and Blue Dots: simply stratified Broad Line Regions}",
      journal = {\mnras},
         year = 2026,
        month = mar,
         note = {submitted},
          doi = {10.48550/arXiv.2603.22277},
       adsurl = {https://ui.adsabs.harvard.edu/abs/2026arXiv260322277S}
}

@ARTICLE{MadauMaiolino2026,
       author = {{Madau}, Piero and {Maiolino}, Roberto},
        title = "{Little Red Dots as Obscured Little Blue Dots: A Super-Eddington Unification Model}",
      journal = {\aap},
         year = 2026,
        month = oct,
       volume = {714},
        pages = {A276},
archivePrefix = {arXiv},
       eprint = {2602.22386},
 primaryClass = {astro-ph.GA},
          doi = {10.1051/0004-6361/202659606},
       adsurl = {https://ui.adsabs.harvard.edu/abs/2026A&A...714A.276M}
}

@ARTICLE{MadauMaiolinoLF2026,
       author = {{Madau}, Piero and {Maiolino}, Roberto},
        title = "{Little Red Dots as Obscured Little Blue Dots: Relative Abundances, Luminosities, and Black Hole Masses}",
      journal = {\aap},
         year = 2026,
        month = oct,
       volume = {715},
        pages = {A15},
archivePrefix = {arXiv},
       eprint = {2605.05074},
 primaryClass = {astro-ph.GA},
          doi = {10.1051/0004-6361/202660815},
       adsurl = {https://ui.adsabs.harvard.edu/abs/2026A&A...715A..15M}
}

@ARTICLE{Zhou2019Inflow,
       author = {{Zhou}, Hongyan and {Shi}, Xiheng and {Yuan}, Weimin and
                 {Hao}, Lei and {Chen}, Xiangjun and others},
        title = "{Fast inflows as the adjacent fuel of supermassive black hole accretion disks in quasars}",
      journal = {Nature},
         year = 2019,
        month = sep,
       volume = {573},
        pages = {83--86},
          doi = {10.1038/s41586-019-1510-y},
archivePrefix = {arXiv},
       eprint = {1909.07582},
 primaryClass = {astro-ph.GA}
}

@ARTICLE{Kocevski2025,
       author = {{Kocevski}, Dale D. and {Finkelstein}, Steven L. and {Barro}, Guillermo and {Taylor}, Anthony J. and {Calabr{\`o}}, Antonello and {Laloux}, Brivael and {Buchner}, Johannes and {Trump}, Jonathan R. and {Leung}, Gene C.~K. and {Yang}, Guang and {Dickinson}, Mark and {P{\'e}rez-Gonz{\'a}lez}, Pablo G. and {Pacucci}, Fabio and {Inayoshi}, Kohei and {Somerville}, Rachel S. and {McGrath}, Elizabeth J. and {Akins}, Hollis B. and {Bagley}, Micaela B. and {Bowler}, Rebecca A.~A. and {Bisigello}, Laura and {Carnall}, Adam and {Casey}, Caitlin M. and {Cheng}, Yingjie and {Cleri}, Nikko J. and {Costantin}, Luca and {Cullen}, Fergus and {Davis}, Kelcey and {Donnan}, Callum T. and {Dunlop}, James S. and {Ellis}, Richard S. and {Ferguson}, Henry C. and {Fujimoto}, Seiji and {Fontana}, Adriano and {Giavalisco}, Mauro and {Grazian}, Andrea and {Grogin}, Norman A. and {Hathi}, Nimish P. and {Hirschmann}, Michaela and {Huertas-Company}, Marc and {Holwerda}, Benne W. and {Illingworth}, Garth and {Juneau}, St{\'e}phanie and {Kartaltepe}, Jeyhan S. and {Koekemoer}, Anton M. and {Li}, Wenxiu and {Lucas}, Ray A. and {Magee}, Dan and {Mason}, Charlotte and {McLeod}, Derek J. and {McLure}, Ross J. and {Napolitano}, Lorenzo and {Papovich}, Casey and {Pirzkal}, Nor and {Rodighiero}, Giulia and {Santini}, Paola and {Wilkins}, Stephen M. and {Yung}, L.~Y. Aaron},
        title = "{The Rise of Faint, Red Active Galactic Nuclei at z > 4: A Sample of Little Red Dots in the JWST Extragalactic Legacy Fields}",
      journal = {\apj},
         year = 2025,
        month = jun,
       volume = {986},
       number = {2},
          eid = {126},
        pages = {126},
          doi = {10.3847/1538-4357/adbc7d},
archivePrefix = {arXiv},
       eprint = {2404.03576},
 primaryClass = {astro-ph.GA},
       adsurl = {https://ui.adsabs.harvard.edu/abs/2025ApJ...986..126K}
}

@ARTICLE{Juod2024JADES,
       author = {{Juod{\v{z}}balis}, Ignas and {Ji}, Xihan and {Maiolino}, Roberto and {D'Eugenio}, Francesco and {Scholtz}, Jan and {Risaliti}, Guido and {Fabian}, Andrew C. and {Mazzolari}, Giovanni and {Gilli}, Roberto and {Prandoni}, Isabella and {Arribas}, Santiago and {Bunker}, Andrew J. and {Carniani}, Stefano and {Charlot}, St{\'e}phane and {Curtis-Lake}, Emma and {de Graaff}, Anna and {Hainline}, Kevin and {Parlanti}, Eleonora and {Perna}, Michele and {P{\'e}rez-Gonz{\'a}lez}, Pablo G. and {Robertson}, Brant and {Tacchella}, Sandro and {{\"U}bler}, Hannah and {Williams}, Christina C. and {Willott}, Chris and {Witstok}, Joris},
        title = "{JADES - the Rosetta stone of JWST-discovered AGN: deciphering the intriguing nature of early AGN}",
      journal = {\mnras},
         year = 2024,
        month = nov,
       volume = {535},
       number = {1},
        pages = {853-873},
          doi = {10.1093/mnras/stae2367},
archivePrefix = {arXiv},
       eprint = {2407.08643},
 primaryClass = {astro-ph.GA},
       adsurl = {https://ui.adsabs.harvard.edu/abs/2024MNRAS.535..853J}
}

@ARTICLE{Lin2026,
       author = {{Lin}, Xiaojing and {Fan}, Xiaohui and {Cai}, Zheng and {Bian}, Fuyan and {Liu}, Hanpu and {Sun}, Fengwu and {Ma}, Yilun and {Greene}, Jenny E. and {Strauss}, Michael A. and {Green}, Richard and {Lyu}, Jianwei and {Champagne}, Jaclyn B. and {Goulding}, Andy D. and {Inayoshi}, Kohei and {Jin}, Xiangyu and {Leung}, Gene C.~K. and {Li}, Mingyu and {Liu}, Weizhe and {Liu}, Yichen and {Mao}, Junjie and {Pudoka}, Maria Anne and {Tee}, Wei Leong and {Wang}, Ben and {Wang}, Feige and {Wu}, Yunjing and {Yang}, Jinyi and {Zhang}, Haowen and {Zhu}, Yongda},
        title = "{The Discovery of Little Red Dots in the Local Universe: Signatures of Cool Gas Envelopes}",
      journal = {\apj},
         year = 2026,
        month = feb,
       volume = {997},
       number = {2},
          eid = {364},
        pages = {364},
          doi = {10.3847/1538-4357/ae2bdf},
archivePrefix = {arXiv},
       eprint = {2507.10659},
 primaryClass = {astro-ph.GA},
       adsurl = {https://ui.adsabs.harvard.edu/abs/2026ApJ...997..364L}
}

@ARTICLE{Greene2024,
       author = {{Greene}, Jenny E. and {Labbe}, Ivo and {Goulding}, Andy D. and {Furtak}, Lukas J. and {Chemerynska}, Iryna and {Kokorev}, Vasily and {Dayal}, Pratika and {Volonteri}, Marta and {Williams}, Christina C. and {Wang}, Bingjie and {Setton}, David J. and {Burgasser}, Adam J. and {Bezanson}, Rachel and {Atek}, Hakim and {Brammer}, Gabriel and {Cutler}, Sam E. and {Feldmann}, Robert and {Fujimoto}, Seiji and {Glazebrook}, Karl and {de Graaff}, Anna and {Khullar}, Gourav and {Leja}, Joel and {Marchesini}, Danilo and {Maseda}, Michael V. and {Matthee}, Jorryt and {Miller}, Tim B. and {Naidu}, Rohan P. and {Nanayakkara}, Themiya and {Oesch}, Pascal A. and {Pan}, Richard and {Papovich}, Casey and {Price}, Sedona H. and {van Dokkum}, Pieter and {Weaver}, John R. and {Whitaker}, Katherine E. and {Zitrin}, Adi},
        title = "{UNCOVER Spectroscopy Confirms the Surprising Ubiquity of Active Galactic Nuclei in Red Sources at z > 5}",
      journal = {\apj},
         year = 2024,
        month = mar,
       volume = {964},
       number = {1},
          eid = {39},
        pages = {39},
          doi = {10.3847/1538-4357/ad1e5f},
archivePrefix = {arXiv},
       eprint = {2309.05714},
 primaryClass = {astro-ph.GA},
       adsurl = {https://ui.adsabs.harvard.edu/abs/2024ApJ...964...39G}
}

@ARTICLE{Greene2026CaT,
       author = {{Greene}, Jenny E. and {Liu}, Hanpu and {Hviding}, Raphael E. and {Naidu}, Rohan P. and {Setton}, David J. and {Wang}, Bingjie and {Labbe}, Ivo and {Kokorev}, Vasily and {Lin}, Xiaojing and {Akins}, Hollis B. and others},
        title = "{Calcium Triplet Absorption is Common around Little Red Dots}",
      journal = {arXiv e-prints},
         year = 2026,
        month = sep,
          eid = {arXiv:2609.25349},
        pages = {arXiv:2609.25349},
archivePrefix = {arXiv},
       eprint = {2609.25349},
 primaryClass = {astro-ph.GA}
}

@ARTICLE{ChaffeeWhite1982,
       author = {{Chaffee}, F.~H. and {White}, R.~E.},
        title = "{A Survey of Interstellar Neutral Potassium. I. Abundances and Physical Conditions in Clouds toward 188 Early-Type Stars}",
      journal = {\apjs},
         year = 1982,
        month = nov,
       volume = {50},
        pages = {169--198},
          doi = {10.1086/190824}
}

@ARTICLE{Chatzikos2023,
       author = {{Chatzikos}, M. and {Bianchi}, S. and {Camilloni}, F. and {Chakraborty}, P. and {Gunasekera}, C.~M. and {Guzm{\'a}n}, F. and {Milby}, J.~S. and {Sarkar}, A. and {Shaw}, G. and {van Hoof}, P.~A.~M. and {Ferland}, G.~J.},
        title = "{The 2023 Release of Cloudy}",
      journal = {\rmxaa},
         year = 2023,
        month = oct,
       volume = {59},
        pages = {327-343},
          doi = {10.22201/ia.01851101p.2023.59.02.12},
archivePrefix = {arXiv},
       eprint = {2308.06396},
 primaryClass = {astro-ph.GA},
       adsurl = {https://ui.adsabs.harvard.edu/abs/2023RMxAA..59..327C}
}

@ARTICLE{Gunasekera2025,
       author = {{Gunasekera}, Chamani M. and {van Hoof}, Peter A.~M. and
                 {Dehghanian}, Maryam and {Chakraborty}, Priyanka and
                 {Shaw}, Gargi and {Bianchi}, Stefano and {Chatzikos}, Marios and
                 {Tsujimoto}, Masahiro and {Ferland}, Gary J.},
        title = "{The 2025 Release of Cloudy}",
      journal = {\rmxaa},
         year = 2025,
       volume = {61},
        pages = {120--133},
          doi = {10.22201/ia.01851101p.2025.61.03.01},
archivePrefix = {arXiv},
       eprint = {2508.01102},
 primaryClass = {astro-ph.IM}
}

@ARTICLE{Hall2007,
       author = {{Hall}, Patrick B.},
        title = "{A Balmer-line Broad Absorption Line Quasar}",
      journal = {\aj},
         year = 2007,
        month = mar,
       volume = {133},
        pages = {1271--1274},
          doi = {10.1086/511272},
archivePrefix = {arXiv},
       eprint = {astro-ph/0611922}
}

@ARTICLE{Yan2025,
       author = {{Yan}, Zu and {Inayoshi}, Kohei and {Chen}, Kejian and {Guo}, Jingsong},
        title = "{Balmer Transition Signatures from Gas-Enshrouded, Dust-Poor Active Galactic Nuclei}",
      journal = {arXiv e-prints},
         year = 2025,
        month = dec,
          eid = {arXiv:2512.11050},
        pages = {arXiv:2512.11050},
          doi = {10.48550/arXiv.2512.11050},
archivePrefix = {arXiv},
       eprint = {2512.11050},
 primaryClass = {astro-ph.GA},
       adsurl = {https://ui.adsabs.harvard.edu/abs/2025arXiv251211050Y}
}

@ARTICLE{Netzer2020,
       author = {{Netzer}, Hagai},
        title = "{Testing broad-line region models with reverberation mapping}",
      journal = {\mnras},
         year = 2020,
        month = may,
       volume = {494},
       number = {2},
        pages = {1611-1621},
          doi = {10.1093/mnras/staa767},
archivePrefix = {arXiv},
       eprint = {2003.07660},
 primaryClass = {astro-ph.GA},
       adsurl = {https://ui.adsabs.harvard.edu/abs/2020MNRAS.494.1611N}
}

@ARTICLE{Bentz2013,
  author       = {Bentz, Misty C. and Denney, Kelly D. and Grier, Catherine J. and Barth, Aaron J. and Peterson, Bradley M. and Vestergaard, Marianne and Bennert, Vardha N. and Canalizo, Gabriela and De Rosa, Gisella and Filippenko, Alexei V. and Greene, Jenny E. and Li, Weidong and Malkan, Matthew A. and Pogge, Richard W. and Stern, Daniel and Treu, Tommaso and Woo, Jong-Hak},
  title        = {The Low-Luminosity End of the Radius-Luminosity Relationship for Active Galactic Nuclei},
  journal      = {\apj},
  year         = {2013},
  volume       = {767},
  pages        = {149},
  doi          = {10.1088/0004-637X/767/2/149},
  adsurl       = {https://ui.adsabs.harvard.edu/abs/2013ApJ...767..149B}
}

@ARTICLE{Cardelli1989,
       author = {{Cardelli}, Jason A. and {Clayton}, Geoffrey C. and {Mathis}, John S.},
        title = "{The Relationship between Infrared, Optical, and Ultraviolet Extinction}",
      journal = {\apj},
         year = 1989,
        month = oct,
       volume = {345},
        pages = {245},
          doi = {10.1086/167900},
       adsurl = {https://ui.adsabs.harvard.edu/abs/1989ApJ...345..245C}
}

@ARTICLE{BruzualCharlot2003,
       author = {{Bruzual}, Gustavo and {Charlot}, St{\'e}phane},
        title = "{Stellar population synthesis at the resolution of 2003}",
      journal = {\mnras},
         year = 2003,
        month = oct,
       volume = {344},
       number = {4},
        pages = {1000-1028},
          doi = {10.1046/j.1365-8711.2003.06897.x},
archivePrefix = {arXiv},
       eprint = {astro-ph/0309134},
 primaryClass = {astro-ph},
}

@ARTICLE{Chabrier2003,
       author = {{Chabrier}, Gilles},
        title = "{Galactic Stellar and Substellar Initial Mass Function}",
      journal = {\pasp},
         year = 2003,
        month = jul,
       volume = {115},
       number = {809},
        pages = {763-795},
          doi = {10.1086/376392},
archivePrefix = {arXiv},
       eprint = {astro-ph/0304382},
 primaryClass = {astro-ph},
}

@ARTICLE{Zafar2015,
       author = {{Zafar}, Tayyaba and {M{\o}ller}, Palle and {Watson}, Darach and {Fynbo}, Johan P.~U. and {Krogager}, Jens-Kristian and {Zafar}, Nosheen and {Saturni}, Francesco G. and {Geier}, Stefan and {Venemans}, Bram P.},
        title = "{Extinction curve template for intrinsically reddened quasars}",
      journal = {\aap},
         year = 2015,
        month = dec,
       volume = {584},
          eid = {A100},
        pages = {A100},
          doi = {10.1051/0004-6361/201526570},
archivePrefix = {arXiv},
       eprint = {1510.01708},
 primaryClass = {astro-ph.GA},
}

@ARTICLE{Madau2024,
       author = {{Madau}, Piero and {Haardt}, Francesco},
        title = "{X-Ray Weak Active Galactic Nuclei from Super-Eddington Accretion onto Infant Black Holes}",
      journal = {\apjl},
         year = 2024,
        month = dec,
       volume = {976},
       number = {2},
          eid = {L24},
        pages = {L24},
          doi = {10.3847/2041-8213/ad90e1},
       adsurl = {https://ui.adsabs.harvard.edu/abs/2024ApJ...976L..24M}
}

@ARTICLE{Torralba2026b,
       author = {{Torralba}, Alberto and {Matthee}, Jorryt and {Pezzulli}, Gabriele and {Naidu}, Rohan P. and {Ishikawa}, Yuzo and {Brammer}, Gabriel B. and {Chang}, Seok-Jun and {Chisholm}, John and {de Graaff}, Anna and {D'Eugenio}, Francesco and {Di Cesare}, Claudia and {Eilers}, Anna-Christina and {Greene}, Jenny E. and {Gronke}, Max and {Iani}, Edoardo and {Kokorev}, Vasily and {Kotiwale}, Gauri and {Kramarenko}, Ivan and {Ma}, Yilun and {Mascia}, Sara and {Navarrete}, Benjam{\'\i}n and {Nelson}, Erica and {Oesch}, Pascal and {Simcoe}, Robert A. and {Wuyts}, Stijn},
        title = "{The warm outer layer of a little red dot as the source of [Fe II] and collisional Balmer lines with scattering wings}",
      journal = {\aap},
         year = 2026,
        month = feb,
       volume = {707},
          eid = {A75},
        pages = {A75},
          doi = {10.1051/0004-6361/202557537},
archivePrefix = {arXiv},
       eprint = {2510.00103},
 primaryClass = {astro-ph.GA},
       adsurl = {https://ui.adsabs.harvard.edu/abs/2026A&A...707A..75T}
}

@ARTICLE{Trinca2026,
       author = {{Trinca}, Alessandro and {Lupi}, Alessandro and {Haardt}, Francesco and {Madau}, Piero},
        title = "{You can't see me: Super-Eddington growth hindering X-ray detection in high-z broad-line active galactic nuclei}",
      journal = {\aap},
         year = 2026,
        month = jun,
       volume = {710},
          eid = {A289},
        pages = {A289},
          doi = {10.1051/0004-6361/202659544},
       adsurl = {https://ui.adsabs.harvard.edu/abs/2026A&A...710A.289T}
}

@ARTICLE{Tortosa2026,
       author = {{Tortosa}, Alessia and others},
        title = "{Widespread steep X-ray spectra of luminous $z=5.85$--$7.64$ quasars and their implications for future X-ray surveys}",
      journal = {\aap},
         year = 2026,
         note = {in press},
archivePrefix = {arXiv},
       eprint = {2610.10522},
 primaryClass = {astro-ph.HE},
          doi = {10.48550/arXiv.2610.10522},
       adsurl = {https://ui.adsabs.harvard.edu/abs/2026arXiv261010522T}
}

@ARTICLE{Madau2026,
  author   = {{Madau}, Piero},
  title    = {Chasing the light: Shadowing, collimation, and the super-Eddington growth of infant black holes in JWST broad-line AGNs},
  journal  = {\aap},
  year     = {2026},
  volume   = {708},
  pages    = {A116},
  doi      = {10.1051/0004-6361/202659244},
  adsurl   = {https://ui.adsabs.harvard.edu/abs/2026A&A...708A.116M}
}

@ARTICLE{Matthee2024,
       author = {{Matthee}, Jorryt and {Naidu}, Rohan P. and {Brammer}, Gabriel and {Chisholm}, John and {Eilers}, Anna-Christina and {Goulding}, Andy and {Greene}, Jenny and {Kashino}, Daichi and {Labbe}, Ivo and {Lilly}, Simon J. and {Mackenzie}, Ruari and {Oesch}, Pascal A. and {Weibel}, Andrea and {Wuyts}, Stijn and {Xiao}, Mengyuan and {Bordoloi}, Rongmon and {Bouwens}, Rychard and {van Dokkum}, Pieter and {Illingworth}, Garth and {Kramarenko}, Ivan and {Maseda}, Michael V. and {Mason}, Charlotte and {Meyer}, Romain A. and {Nelson}, Erica J. and {Reddy}, Naveen A. and {Shivaei}, Irene and {Simcoe}, Robert A. and {Yue}, Minghao},
        title = "{Little Red Dots: An Abundant Population of Faint Active Galactic Nuclei at z {\ensuremath{\sim}} 5 Revealed by the EIGER and FRESCO JWST Surveys}",
      journal = {\apj},
         year = 2024,
        month = mar,
       volume = {963},
       number = {2},
          eid = {129},
        pages = {129},
          doi = {10.3847/1538-4357/ad2345},
archivePrefix = {arXiv},
       eprint = {2306.05448},
 primaryClass = {astro-ph.GA},
       adsurl = {https://ui.adsabs.harvard.edu/abs/2024ApJ...963..129M}
}

@ARTICLE{Panda2021,
       author = {{Panda}, Swayamtrupta},
        title = "{The CaFe project: Optical Fe II and near-infrared Ca II triplet emission in active galaxies: simulated EWs and the co-dependence of cloud size and metal content}",
      journal = {\aap},
         year = 2021,
        month = jun,
       volume = {650},
          eid = {A154},
        pages = {A154},
          doi = {10.1051/0004-6361/202140393},
archivePrefix = {arXiv},
       eprint = {2004.13113},
 primaryClass = {astro-ph.GA},
       adsurl = {https://ui.adsabs.harvard.edu/abs/2021A&A...650A.154P}
}

@ARTICLE{Matsuoka2007,
       author = {{Matsuoka}, Yoshiki and {Oyabu}, Shinki and {Tsuzuki}, Yumihiko and {Kawara}, Kimiaki},
        title = "{Observations of O I and Ca II Emission Lines in Quasars: Implications for the Site of Fe II Line Emission}",
      journal = {\apj},
         year = 2007,
        month = jul,
       volume = {663},
       number = {2},
        pages = {781--798},
          doi = {10.1086/518399},
archivePrefix = {arXiv},
       eprint = {astro-ph/0703659}
}

@ARTICLE{Leighly2014,
       author = {{Leighly}, Karen M. and {Terndrup}, Donald M. and {Baron}, Eddie and {Lucy}, Adrian B. and {Dietrich}, Matthias and {Gallagher}, Sarah C.},
        title = "{Evidence for Active Galactic Nucleus Feedback in the Broad Absorption Lines and Reddening of Mrk 231}",
      journal = {\apj},
         year = 2014,
        month = jun,
       volume = {788},
       number = {2},
          eid = {123},
        pages = {123},
          doi = {10.1088/0004-637X/788/2/123},
archivePrefix = {arXiv},
       eprint = {1404.6795}
}

@ARTICLE{Leighly2025Balmer,
       author = {{Leighly}, Karen M. and {Gallagher}, Sarah C. and
                 {Choi}, Hyunseop and {Terndrup}, Donald M. and
                 {Voelker}, Julianna R. and {Richards}, Gordon T. and
                 {Morabito}, Leah K.},
        title = "{Balmer Absorption in Iron Low-Ionization Broad Absorption Line Quasars}",
      journal = {\apj\ },
         year = 2025,
         note = {in press},
          doi = {10.48550/arXiv.2509.07611},
archivePrefix = {arXiv},
       eprint = {2509.07611},
 primaryClass = {astro-ph.GA}
}

@ARTICLE{Tripodi2025a,
       author = {{Tripodi}, Roberta and {Brada{\v{c}}}, Maru{\v{s}}a and {D'Eugenio}, Francesco and {Martis}, Nicholas and {Rihtar{\v{s}}i{\v{c}}}, Gregor and {Willott}, Chris and {Pentericci}, Laura and {Moreschini}, Bianca and {Markevitch}, Maxim and {Asada}, Yoshihisa and {Calabr{\'o}}, Antonello and {Desprez}, Guillaume and {Felicioni}, Giordano and {Gaspar}, Gaia and {Gonzalez}, Anthony H. and {Harshan}, Anishya and {Ji}, Xihan and {Jude{\v{z}}}, Jon and {Lemaux}, Brian C. and {Marconi}, Alessandro and {Markov}, Vladan and {Merida}, Rosa M. and {Napolitano}, Lorenzo and {Noirot}, Ga{\"e}l and {Parente}, Massimiliano and {Peter}, Annika H.~G. and {Robbins}, Luke and {Robertson}, Andrew and {Sarrouh}, Ghassan T.~E. and {Sawicki}, Marcin},
        title = "{A Deep Dive down the Broad-line Region: Permitted O I, Ca II, and Fe II Emission in an Active Galactic Nucleus Little Red Dot at z = 5.3}",
      journal = {\apjl},
         year = 2025,
        month = nov,
       volume = {994},
       number = {1},
          eid = {L6},
        pages = {L6},
          doi = {10.3847/2041-8213/ae13a9},
archivePrefix = {arXiv},
       eprint = {2507.20684},
 primaryClass = {astro-ph.GA},
       adsurl = {https://ui.adsabs.harvard.edu/abs/2025ApJ...994L...6T}
}

@ARTICLE{Ji2025z01,
       author = {{Ji}, Xihan and {D'Eugenio}, Francesco and {Juod{\v{z}}balis}, Ignas and {Walton}, Dominic J. and {Fabian}, Andrew C. and {Maiolino}, Roberto and {Almeida}, Cristina Ramos and {Pulido}, Jose A. Acosta and {Belokurov}, Vasily A. and {Isobe}, Yuki and {Jones}, Gareth and {Maraston}, Claudia and {Scholtz}, Jan and {Simmonds}, Charlotte and {Tacchella}, Sandro and {Terlevich}, Elena and {Terlevich}, Roberto},
        title = "{Lord of LRDs: Insights into a 'Little Red Dot' with a low-ionization spectrum at z = 0.1}",
      journal = {\mnras},
         year = 2026,
        month = jan,
       volume = {545},
       number = {3},
          eid = {staf2235},
        pages = {staf2235},
          doi = {10.1093/mnras/staf2235},
archivePrefix = {arXiv},
       eprint = {2507.23774},
 primaryClass = {astro-ph.GA},
       adsurl = {https://ui.adsabs.harvard.edu/abs/2025MNRAS.tmp.2102J}
}

@ARTICLE{Ji2026b,
       author = {{Ji}, Xihan and {Walton}, Dominic J. and {D'Eugenio}, Francesco and {Juod{\v{z}}balis}, Ignas and {Belokurov}, Vasily and {Fabian}, Andrew C. and {Maiolino}, Roberto and {Huang}, Jiamu and {Lin}, Douglas N.~C.},
        title = "{Gone with the wind? Potential gas inflow in the broad-line region of a QSO at z = 0.287 traced by time-varying Balmer absorption}",
      journal = {arXiv e-prints},
         year = 2026,
        month = sep,
          eid = {arXiv:2609.25224},
        pages = {arXiv:2609.25224},
          doi = {10.48550/arXiv.2609.25224},
archivePrefix = {arXiv},
       eprint = {2609.25224},
 primaryClass = {astro-ph.GA}
}

@ARTICLE{Wang2026Balmer,
       author = {{Wang}, Bingjie and {Greene}, Jenny E. and {Liu}, Hanpu and
                 {Kaaz}, Nicholas and {Brammer}, Gabriel B. and {Hviding}, Raphael E. and
                 {Labb{\'e}}, Ivo and {Leja}, Joel and {Matthee}, Jorryt and
                 {Naidu}, Rohan P. and others},
        title = "{Balmer Absorption Series and Broad Metal Lines in Two Luminous Little Red Dots}",
      journal = {arXiv e-prints},
         year = 2026,
        month = sep,
          eid = {arXiv:2609.22623},
        pages = {arXiv:2609.22623},
          doi = {10.48550/arXiv.2609.22623},
archivePrefix = {arXiv},
       eprint = {2609.22623},
 primaryClass = {astro-ph.GA}
}

@ARTICLE{Inayoshi2025,
       author = {{Inayoshi}, Kohei and {Maiolino}, Roberto},
        title = "{Extremely Dense Gas around Little Red Dots and High-redshift Active Galactic Nuclei: A Nonstellar Origin of the Balmer Break and Absorption Features}",
      journal = {\apjl},
         year = 2025,
        month = feb,
       volume = {980},
       number = {2},
          eid = {L27},
        pages = {L27},
          doi = {10.3847/2041-8213/adaebd},
archivePrefix = {arXiv},
       eprint = {2409.07805},
 primaryClass = {astro-ph.GA},
       adsurl = {https://ui.adsabs.harvard.edu/abs/2025ApJ...980L..27I}
}

@ARTICLE{IzotovThuan2008,
       author = {{Izotov}, Yuri I. and {Thuan}, Trinh X.},
        title = "{Active Galactic Nuclei in Four Metal-poor Dwarf Emission-line Galaxies}",
      journal = {\apj},
         year = 2008,
        month = nov,
       volume = {687},
        pages = {133-140},
          doi = {10.1086/591660},
archivePrefix = {arXiv},
       eprint = {0807.2029}
}

@ARTICLE{Simmonds2016,
       author = {{Simmonds}, Charlotte and {Bauer}, Franz E. and {Thuan}, Trinh X. and {Izotov}, Yuri I. and {Stern}, Daniel and {Harrison}, Fiona A.},
        title = "{Do Some AGN Lack X-ray Emission?}",
      journal = {\aap},
         year = 2016,
        month = dec,
       volume = {596},
          eid = {A64},
        pages = {A64},
          doi = {10.1051/0004-6361/201629310},
archivePrefix = {arXiv},
       eprint = {1609.07619}
}

@ARTICLE{Burke2021,
       author = {{Burke}, Colin J. and {Liu}, Xin and {Chen}, Yu-Ching and {Shen}, Yue and {Guo}, Hengxiao},
        title = "{On the AGN Nature of Broad Balmer Emission in Four Low-redshift Metal-poor Galaxies}",
      journal = {\mnras},
         year = 2021,
        month = jun,
       volume = {504},
        pages = {543-550},
          doi = {10.1093/mnras/stab912}
}

@ARTICLE{Burke2026Quiet,
       author = {{Burke}, Colin J. and {Stone}, Zachary and {Shen}, Yue and {Jiang}, Yan-Fei},
        title = "{Too Quiet for Comfort: Local Little Red Dots Lack Variability over Decades}",
      journal = {\apj},
         year = 2026,
        month = jun,
       volume = {1004},
       number = {1},
          eid = {16},
        pages = {16},
          doi = {10.3847/1538-4357/ae6b89},
archivePrefix = {arXiv},
       eprint = {2511.16082}
}

@ARTICLE{Liu2026Atmospheres,
       author = {{Liu}, Hanpu and {Jiang}, Yan-Fei and {Quataert}, Eliot and {Greene}, Jenny E. and {Ma}, Yilun and {Lin}, Xiaojing},
        title = "{Synthetic Spectral Library of Optically Thick Atmospheres for Little Red Dots}",
      journal = {arXiv e-prints},
         year = 2026,
        month = mar,
        pages = {arXiv:2603.02317},
archivePrefix = {arXiv},
       eprint = {2603.02317},
 primaryClass = {astro-ph.GA}
}

@ARTICLE{Liu2026Balmer,
       author = {{Liu}, Hanpu and {Quataert}, Eliot and {Wang}, Bingjie and
                 {Greene}, Jenny E. and {Jiang}, Yan-Fei},
        title = "{The Balmer Absorber Strength Problem of Little Red Dots}",
      journal = {arXiv e-prints},
         year = 2026,
        month = oct,
          eid = {arXiv:2610.28720},
        pages = {arXiv:2610.28720},
          doi = {10.48550/arXiv.2610.28720},
archivePrefix = {arXiv},
       eprint = {2610.28720},
 primaryClass = {astro-ph.GA}
}

@ARTICLE{Shangguan2026,
       author = {{Shangguan}, Jinyi and {Chen}, Chang-Hao and {Ho}, Luis C. and {Liao}, Jiwei and {Liu}, Yanqing and {Wu}, Chengzhou and {Li}, Ruancun and {Inayoshi}, Kohei and {Jiang}, Linhua},
        title = "{The Extreme Rarity and Physical Properties of Low-redshift AGNs with Balmer Absorption}",
      journal = {arXiv e-prints},
         year = 2026,
        month = jun,
        pages = {arXiv:2606.04712},
archivePrefix = {arXiv},
       eprint = {2606.04712},
 primaryClass = {astro-ph.GA}
}

@ARTICLE{DEugenio2026Polarization,
       author = {{D'Eugenio}, Francesco and {Pezzulli}, Gabriele and {Maiolino}, Roberto and {Marconi}, Alessandro and {Ji}, Xihan and {Ramos Almeida}, Cristina and {Ferrara}, Andrea and {Madau}, Piero and {Lin}, Xiaojing and others},
        title = "{Misaligned or Chaotic? A Strong Break of Axial Symmetry in the Local LRD J1025 Revealed with VLT/FORS2 Spectropolarimetry}",
      journal = {arXiv e-prints},
         year = 2026,
        month = jul,
        pages = {arXiv:2607.28715},
archivePrefix = {arXiv},
       eprint = {2607.28715},
 primaryClass = {astro-ph.GA}
}

@ARTICLE{MadauMaiolino_POL_2026,
       author = {{Madau}, Piero and {Maiolino}, Roberto and {D'Eugenio}, Francesco},
        title = "{Connecting the Little Dots in Polarized Light}",
      journal = {\aap},
         year = 2026,
        month = sep,
         note = {submitted},
archivePrefix = {arXiv},
       eprint = {2609.06239},
 primaryClass = {astro-ph.GA},
          doi = {10.48550/arXiv.2609.06239},
       adsurl = {https://ui.adsabs.harvard.edu/abs/2026arXiv260906239M}
}

@ARTICLE{Madau2026Wings,
       author = {{Madau}, Piero and {Maiolino}, Roberto and {Scholtz}, Jan and {D'Eugenio}, Francesco},
        title = "{Wings of little dots: Exponential broad lines from a stratified broad-line region}",
      journal = {\aap},
         year = 2026,
        month = oct,
       volume = {714},
        pages = {A186},
archivePrefix = {arXiv},
       eprint = {2604.04216},
 primaryClass = {astro-ph.GA},
          doi = {10.1051/0004-6361/202660251},
       adsurl = {https://ui.adsabs.harvard.edu/abs/2026A&A...714A.186M}
}

@ARTICLE{Kiyota2026,
       author = {{Kiyota}, Tomokazu and {Ouchi}, Masami and {Yanagisawa}, Hiroto and {Ando}, Makoto and {Harikane}, Yuichi and {Kageura}, Yuta and {Nakane}, Minami and {Nakazato}, Yurina and {Ono}, Yoshiaki and {Takeda}, Yui},
        title = "{ATLAS. III. Dust Around Little Red Dots: Hydrogen Line Ratios beyond Dust-free Non-Case B Models}",
      journal = {arXiv e-prints},
         year = 2026,
        month = aug,
        pages = {arXiv:2608.10832},
archivePrefix = {arXiv},
       eprint = {2608.10832},
 primaryClass = {astro-ph.GA}
}

@ARTICLE{Bentz2010,
       author = {{Bentz}, Misty C. and {Walsh}, Jonelle L. and {Barth}, Aaron J. and {Yoshii}, Yuzuru and {Woo}, Jong-Hak and {Wang}, Xiaofeng and {Treu}, Tommaso and {Thornton}, Carol E. and {Street}, Rachel A. and {Steele}, Thea N. and {Silverman}, Jeffrey M. and {Serduke}, Frank J.~D. and {Sakata}, Yu and {Minezaki}, Takeo and {Malkan}, Matthew A. and {Li}, Weidong and {Lee}, Nicholas and {Hiner}, Kyle D. and {Hidas}, Marton G. and {Greene}, Jenny E. and {Gates}, Elinor L. and {Ganeshalingam}, Mohan and {Filippenko}, Alexei V. and {Canalizo}, Gabriela and {Bennert}, Vardha Nicola and {Baliber}, Nairn},
        title = "{The Lick AGN Monitoring Project: Reverberation Mapping of Optical Hydrogen and Helium Recombination Lines}",
      journal = {ApJ},
         year = 2010,
        month = jun,
       volume = {716},
        pages = {993-1011},
          doi = {10.1088/0004-637X/716/2/993},
archivePrefix = {arXiv},
       eprint = {1004.2922}
}

@ARTICLE{Feng2021,
       author = {{Feng}, Hai-Cheng and {Liu}, H.~T. and {Bai}, J.~M. and {Yang}, Zi-Xu and {Hu}, Chen and {Li}, Sha-Sha and {Yang}, Sen and {Lu}, Kai-Xing and {Xiao}, Ming},
        title = "{Velocity-resolved Reverberation Mapping of Changing-look AGN NGC 2617}",
      journal = {ApJ},
         year = 2021,
        month = may,
       volume = {912},
       number = {2},
          eid = {92},
        pages = {92},
          doi = {10.3847/1538-4357/abefe0},
archivePrefix = {arXiv},
       eprint = {2103.03508}
}

@ARTICLE{Wilms2000,
       author = {{Wilms}, J. and {Allen}, A. and {McCray}, R.},
        title = "{On the Absorption of X-Rays in the Interstellar Medium}",
      journal = {\apj},
         year = 2000,
        month = oct,
       volume = {542},
       number = {2},
        pages = {914-924},
          doi = {10.1086/317016}
}

@ARTICLE{Just2007,
       author = {{Just}, D.~W. and {Brandt}, W.~N. and {Shemmer}, O. and
                 {Steffen}, A.~T. and {Schneider}, D.~P. and {Chartas}, G. and
                 {Garmire}, G.~P.},
        title = "{The X-Ray Properties of the Most Luminous Quasars from the Sloan Digital Sky Survey}",
      journal = {\apj},
         year = 2007,
        month = aug,
       volume = {665},
        pages = {1004--1022},
          doi = {10.1086/519990}
}

@ARTICLE{FerlandPersson1989,
       author = {{Ferland}, G.~J. and {Persson}, S.~E.},
        title = "{Implications of Ca II emission for physical conditions in the broad-line region of active galactic nuclei}",
      journal = {ApJ},
         year = 1989,
       volume = {347},
        pages = {656--673},
          doi = {10.1086/168156}
}

@ARTICLE{CollinJoly2000,
       author = {{Collin}, S. and {Joly}, M.},
        title = "{The Fe II problem in NLS1s}",
      journal = {New Astronomy Reviews},
         year = 2000,
       volume = {44},
        pages = {531--537},
          doi = {10.1016/S1387-6473(00)00093-2}
}

@ARTICLE{VeronCetty2004,
       author = {{V{\'e}ron-Cetty}, M.-P. and {Joly}, M. and {V{\'e}ron}, P.},
        title = "{The unusual emission line spectrum of I Zw 1}",
      journal = {A\&A},
         year = 2004,
       volume = {417},
        pages = {515--525},
          doi = {10.1051/0004-6361:20035714}
}

@ARTICLE{Ferland2009,
       author = {{Ferland}, G.~J. and {Fabian}, A.~C. and {Hatch}, N.~A. and
                 {Johnstone}, R.~M. and {Porter}, R.~L. and
                 {van Hoof}, P.~A.~M. and {Williams}, R.~J.~R.},
        title = "{Collisional heating as the origin of filament emission in galaxy clusters}",
      journal = {MNRAS},
         year = 2009,
       volume = {392},
        pages = {1475--1502},
          doi = {10.1111/j.1365-2966.2008.14153.x}
}

@ARTICLE{Vollmer2004,
       author = {{Vollmer}, B. and {Beckert}, T. and {Duschl}, W.~J.},
        title = "{From the circumnuclear disk in the Galactic Center to thick, obscuring tori of AGNs}",
      journal = {A\&A},
         year = 2004,
       volume = {413},
        pages = {949--956},
          doi = {10.1051/0004-6361:20034201}
}

@ARTICLE{KrolikBegelman1988,
       author = {{Krolik}, J.~H. and {Begelman}, M.~C.},
        title = "{Molecular tori in Seyfert galaxies: feeding the monster and hiding it}",
      journal = {ApJ},
         year = 1988,
       volume = {329},
        pages = {702--711},
          doi = {10.1086/166414}
}

@ARTICLE{Vollmer2022,
       author = {{Vollmer}, B. and {Davies}, R.~I. and {Gratier}, P. and
                 {Liz{\'e}e}, T. and {Imanishi}, M. and {Gallimore}, J.~F. and
                 {Impellizzeri}, C.~M.~V. and {Garc{\'i}a-Burillo}, S. and
                 {Le Petit}, F.},
        title = "{From the circumnuclear disk in the Galactic Center to thick, obscuring tori of AGNs: modelling the molecular emission of a parsec-scale torus as found in NGC 1068}",
      journal = {A\&A},
         year = 2022,
       volume = {665},
          eid = {A102},
        pages = {A102},
          doi = {10.1051/0004-6361/202141684}
}

@ARTICLE{Prochaska2011,
       author = {{Prochaska}, J.~X. and {Kasen}, D. and {Rubin}, K.},
        title = "{Simple Models of Metal-line Absorption and Emission from Cool Gas Outflows}",
      journal = {ApJ},
         year = 2011,
       volume = {734},
          eid = {24},
        pages = {24},
          doi = {10.1088/0004-637X/734/1/24}
}

@ARTICLE{RupkeVeilleux2015,
       author = {{Rupke}, D.~S.~N. and {Veilleux}, S.},
        title = "{Spatially Extended Na I D Resonant Emission and Absorption in the Galactic Wind of the Nearby Infrared-luminous Quasar F05189-2524}",
      journal = {ApJ},
         year = 2015,
       volume = {801},
          eid = {126},
        pages = {126},
          doi = {10.1088/0004-637X/801/2/126}
}

@ARTICLE{Chon2026,
       author = {{Chon}, S. and {Hirano}, S. and {Ishiyama}, T. and
                 {Chang}, S.-J. and {Springel}, V.},
        title = "{Overmassive black holes and little red dots naturally form in simulations}",
      journal = {Nature},
         year = 2026,
       volume = {657},
        pages = {621--625},
          doi = {10.1038/s41586-026-10985-8}
}

@ARTICLE{Sanders2025Attenuation,
       author = {{Sanders}, Ryan L. and {Shapley}, Alice E. and {Topping},
                 Michael W. and {Reddy}, Naveen A. and {Berg}, Danielle A. and
                 others},
        title = "{The AURORA Survey: The Nebular Attenuation Curve of a Galaxy
                  at $z=4.41$ from Ultraviolet to Near-infrared Wavelengths}",
      journal = {\apj},
         year = 2025,
       volume = {989},
        pages = {209},
          doi = {10.3847/1538-4357/adf066}
}

@ARTICLE{Reddy2026Attenuation,
       author = {{Reddy}, Naveen A. and {Shapley}, Alice E. and {Sanders},
                 Ryan L. and {Topping}, Michael W. and {Ellis}, Richard S. and
                 {Pettini}, Max and {Brammer}, Gabriel and {Cullen}, Fergus and
                 {F{\"o}rster Schreiber}, N. M. and {Khostovan}, Ali A. and
                 {McLeod}, Derek J. and {McLure}, Ross J. and {Narayanan},
                 Desika and {Oesch}, Pascal A. and {Pahl}, Anthony J. and
                 {Steidel}, Charles C. and {Berg}, Danielle A.},
        title = "{The AURORA Survey: Multiple Balmer and Paschen Emission Lines
                  for Individual Star-forming Galaxies at $z=1.5$--$4.4$. I.
                  A Diversity of Nebular Attenuation Curves and Evidence for
                  Non-unity Dust Covering Fractions}",
      journal = {\apj},
         year = 2026,
       volume = {999},
        pages = {15},
          doi = {10.3847/1538-4357/ae38da}
}

@ARTICLE{CzernyHryniewicz2011,
       author = {{Czerny}, Bo{\.z}ena and {Hryniewicz}, Krzysztof},
        title = "{The origin of the broad line region in active galactic nuclei}",
      journal = {A\&A},
         year = 2011,
       volume = {525},
          eid = {L8},
        pages = {L8},
          doi = {10.1051/0004-6361/201016025}
}

@ARTICLE{NaddafCzerny2022,
       author = {{Naddaf}, Mohammad-Hassan and {Czerny}, Bo{\.z}ena},
        title = "{Radiation pressure on dust explaining the low ionized broad emission lines in active galactic nuclei: Dust as an important driver of line shape}",
      journal = {A\&A},
         year = 2022,
       volume = {663},
          eid = {A77},
        pages = {A77},
          doi = {10.1051/0004-6361/202142806}
}

@ARTICLE{NaddafCzernyZajacek2022,
       author = {{Naddaf}, Mohammad-Hassan and {Czerny}, Bo{\.z}ena and
                 {Zaja{\v{c}}ek}, Michal},
        title = "{The Wind Dynamics of Super-Eddington Sources in FRADO}",
      journal = {Dynamics},
         year = 2022,
       volume = {2},
       number = {3},
        pages = {295--305},
          doi = {10.3390/dynamics2030015}
}

@ARTICLE{Gaskell1982,
       author = {{Gaskell}, C.~M.},
        title = "{A Redshift Difference between High- and Low-Ionization Emission-Line Regions in QSOs: Evidence for Radial Motions}",
      journal = {\apj},
         year = 1982,
        month = dec,
       volume = {263},
        pages = {79--86},
          doi = {10.1086/160481}
}

@ARTICLE{Laor1997,
       author = {{Laor}, Ari and {Jannuzi}, Buell T. and {Green}, Richard F. and
                 {Boroson}, Todd A.},
        title = "{The Ultraviolet Properties of the Narrow-Line Quasar I Zw 1}",
      journal = {\apj},
         year = 1997,
        month = nov,
       volume = {489},
        pages = {656--671},
          doi = {10.1086/304816}
}

@ARTICLE{Shen2016,
       author = {{Shen}, Yue and {Brandt}, W.~N. and {Richards}, G.~T. and
                 {Denney}, K.~D. and {Greene}, J.~E. and {Grier}, C.~J. and
                 {Ho}, L.~C. and {Peterson}, B.~M. and {Petitjean}, P. and
                 {Schneider}, D.~P. and {Tao}, C. and {Trump}, J.~R.},
        title = "{The Sloan Digital Sky Survey Reverberation Mapping Project: Velocity Shifts of Quasar Emission Lines}",
      journal = {\apj},
         year = 2016,
        month = nov,
       volume = {831},
          eid = {7},
        pages = {7},
          doi = {10.3847/0004-637X/831/1/7}
}

@ARTICLE{Meyer2019,
       author = {{Meyer}, Romain A. and {Bosman}, Sarah E.~I. and {Ellis}, Richard S.},
        title = "{New Constraints on Quasar Evolution: Broad-line Velocity Shifts over $1.5\lesssim z\lesssim7.5$}",
      journal = {\mnras},
         year = 2019,
        month = aug,
       volume = {487},
       number = {3},
        pages = {3305--3323},
          doi = {10.1093/mnras/stz1504}
}

@ARTICLE{Gibson2009,
       author = {{Gibson}, R.~R. and {Jiang}, L. and {Brandt}, W.~N. and
                 {Hall}, P.~B. and {Shen}, Y. and {Wu}, J. and {Anderson}, S.~F.
                 and {Schneider}, D.~P. and {Vanden Berk}, D. and
                 {Gallagher}, S.~C. and {Fan}, X. and {York}, D.~G.},
        title = "{A Catalog of Broad Absorption Line Quasars in Sloan Digital Sky Survey Data Release 5}",
      journal = {\apj},
         year = 2009,
        month = feb,
       volume = {692},
       number = {1},
        pages = {758--777},
          doi = {10.1088/0004-637X/692/1/758}
}

@ARTICLE{Hamann2019BAL,
       author = {{Hamann}, Fred and {Herbst}, Hanna and {Paris}, Isabelle and
                 {Capellupo}, Daniel},
        title = "{On the Structure and Energetics of Quasar Broad Absorption-Line Outflows}",
      journal = {\mnras},
         year = 2019,
        month = feb,
       volume = {483},
       number = {2},
        pages = {1808--1828},
          doi = {10.1093/mnras/sty2900},
archivePrefix = {arXiv},
       eprint = {1810.03686},
 primaryClass = {astro-ph.GA}
}

@ARTICLE{Geris2026,
       author = {{Geris}, Sophia and {Maiolino}, Roberto and {Ji}, Xihan and
                 {Risaliti}, Guido and {Lanzuisi}, Giorgio and {D'Eugenio},
                 Francesco and others},
        title = "{Little Red and Blue Dots: AGN-excited narrow lines,
                  Lyman-$\alpha$ emission, and resemblance to standard quasars}",
      journal = {arXiv e-prints},
         year = 2026,
        month = jun,
        pages = {arXiv:2606.21614},
archivePrefix = {arXiv},
       eprint = {2606.21614},
 primaryClass = {astro-ph.GA}
}

@ARTICLE{Li2026SteepExtinction,
       author = {{Li}, Mingyu and {Cai}, Zheng and {Maiolino}, Roberto and
                 {Sun}, Fengwu and {Ji}, Xihan and {Duan}, Qiao and {Emonts},
                 Bjorn H.~C. and {Fan}, Xiaohui and {Juod{\v{z}}balis}, Ignas
                 and {Lin}, Xiaojing and {Liu}, Yixiao and {Tacchella}, Sandro},
        title = "{A Steep-Extinction Quasi-stellar Object at $z=4.6$: JWST
                  Evidence for Abundant Small Dust Grains}",
      journal = {\apjl},
         year = 2026,
       volume = {1008},
        pages = {L22},
archivePrefix = {arXiv},
       eprint = {2606.02685},
 primaryClass = {astro-ph.GA}
}

@ARTICLE{Sun2026SmallGrains,
       author = {{Sun}, Fengwu and {Helton}, Daniel J. and others},
        title = "{JADES: Discovery of Large Reservoirs of Small Dust Grains in
                  the Circumgalactic Medium of Massive Galaxies at $z\sim3.5$
                  through Deep JWST/NIRCam Imaging and Grism Spectroscopy}",
      journal = {arXiv e-prints},
         year = 2026,
        month = jan,
        pages = {arXiv:2601.15961},
archivePrefix = {arXiv},
       eprint = {2601.15961},
 primaryClass = {astro-ph.GA}
}

\begin{appendix}

\section{Polarization as an independent geometric test}
\label{app:polarization}

\begin{figure}[!hb]
 \centering
 \includegraphics[width=0.95\columnwidth]{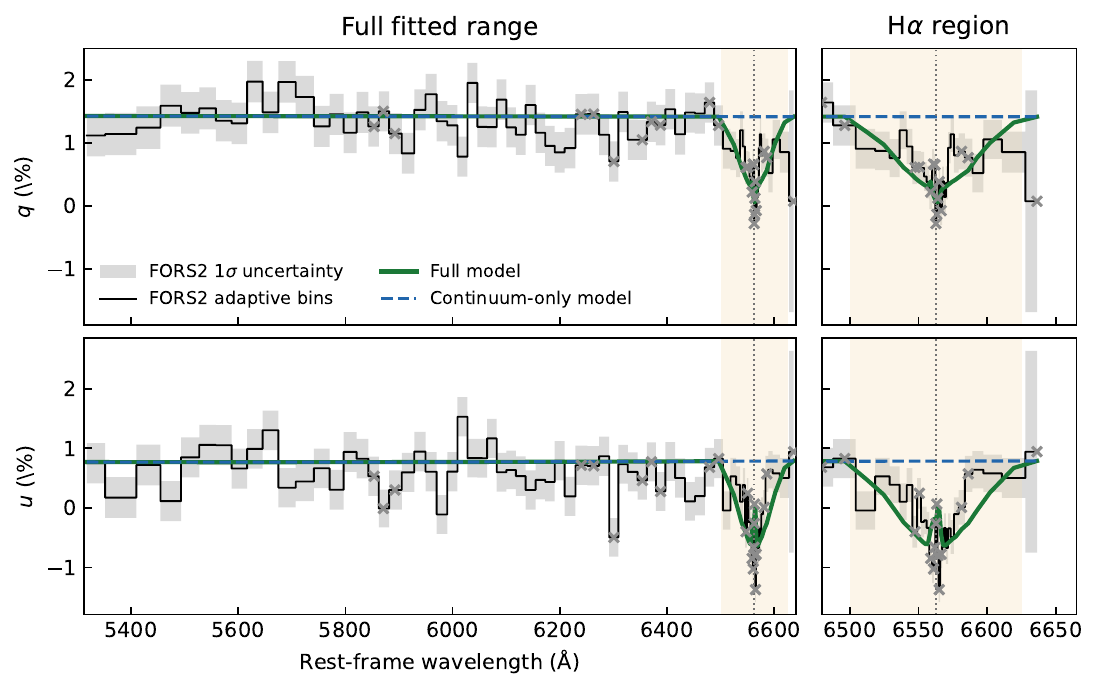}
 \caption{Fit to the normalized FORS2 Stokes spectra
 \citep{DEugenio2026Polarization}. Black and gray show the measurements and
 uncertainties; green is the full model and dashed blue the continuum.
 Gray crosses mark excluded bins, including $\pm200\,\mathrm{km\,s^{-1}}$
 around the narrow H$\alpha$ and [N~{\sc ii}] cores. The direct
 narrow lines are unpolarized, and no additional polarized core component is
 included. Disk parameters were fixed to
 $(\log M_{\rm BH}/M_\odot,\dot m,i)=(7.4,80,73\degr)$; the fit has
 $\chi_\nu^2=1.21$.}
 \label{fig:polarization}
\end{figure}
\citet{DEugenio2026Polarization} measured nearly gray continuum polarization,
$p_{\rm cont}=1.53\pm0.04\,({\rm stat})\pm0.20\,({\rm syst})\%$, broad
H$\alpha$ polarization of 0.58--0.84\%, and a line-to-continuum angle offset
of $-48\degr\pm4\degr\,({\rm stat})\pm12\degr\,({\rm syst})$. Following
\citet{MadauMaiolino_POL_2026}, we combined intrinsic disk polarization,
foreground Serkowski-law dichroism, and dilution by unpolarized direct narrow
H$\alpha$ and [N~{\sc ii}]. All accretion-flow parameters were fixed to the
continuum solution,
$(\log M_{\rm BH}/M_\odot,\dot m,i)=(7.4,80,73\degr)$, which determines the
wavelength-dependent intrinsic disk polarization $p_{\rm disk}(\lambda)$.
Thus, neither inclination nor accretion rate is re-estimated from the
polarization data. We also fixed the Serkowski shape to
$\lambda_{\rm max}=5500$~\AA\ and $K=0.92$; its amplitude and orientation
remain free. That work reproduced the polarization structure in the line
core by adding a polarized narrow-line component. Here we instead
test whether the continuum and broad H$\alpha$ alone are consistent with the
SED-derived geometry, without introducing that component.

The observed intensity spectrum fixes the continuum fraction
$f_{\rm cont}(\lambda)$, while the H$\alpha$ decomposition in
Sect.~\ref{sec:halpha_results} fixes the direct narrow-line fraction
$f_{\rm NL}(\lambda)$. The continuum carries both the disk and dichroic
Stokes vectors, broad H$\alpha$ carries only the dichroic vector, and the
direct narrow lines carry neither. For normalized Stokes parameters $q=Q/I$
and $u=U/I$,
%---------------------------------------------------------------------
\begin{align}
 q_{\rm mod}(\lambda)&=[1-f_{\rm NL}(\lambda)]q_{\rm dic}(\lambda)
       +f_{\rm cont}(\lambda)q_{\rm disk}(\lambda),\nonumber\\
 u_{\rm mod}(\lambda)&=[1-f_{\rm NL}(\lambda)]u_{\rm dic}(\lambda)
       +f_{\rm cont}(\lambda)u_{\rm disk}(\lambda).
 \label{eq:polarization_model}
\end{align}
%---------------------------------------------------------------------
We fitted the directly measured $Q$ and $U$ fluxes with their wavelength-bin
covariance. To isolate this restricted geometric test, we excluded
bins within $200\,\mathrm{km\,s^{-1}}$ of narrow H$\alpha$ and each
[N~{\sc ii}] line. The remaining 57 bins provide 114 Stokes measurements and
111 degrees of freedom for the three fitted parameters $p_{\rm max}$,
$\psi_{\rm disk}$, and $\psi_{\rm dic}$. The result is
%---------------------------------------------------------------------
\begin{align}
 \langle p_{\rm disk}\rangle&=1.99\%, &
 p_{\rm max}&=0.82\pm0.10\%,\nonumber\\
 \psi_{\rm disk}&=26\degr\pm2\degr, &
 \psi_{\rm dic}&=-38\degr\pm3\degr.
 \label{eq:polarization_best}
\end{align}
%---------------------------------------------------------------------
The conditional fit has $\chi^2=134.2$ and $\chi_\nu^2=1.21$. It gives
$p_{\rm cont}=1.62\%$ at $14\degr$ and broad-H$\alpha$ dichroic polarization
of $0.80\%$ at $-38\degr$, hence $\Delta\vartheta=-52\degr$. The agreement
with the observed continuum and broad-line quantities therefore does not
depend on fitting the narrow-line core. The remaining excluded core structure
is not interpreted in this geometric test. Broader-wavelength
spectropolarimetry can further separate nearly gray disk-atmosphere
polarization from wavelength-dependent dichroism.

\section{Supplementary line diagnostics}
\label{app:supplementary_lines}

\subsection{He~{\sc i} $\lambda10833$ and the low-column outflowing skin}
\label{app:hei_absorption}

\citet{Lin2026} measured a P~Cygni-like He~{\sc i} $\lambda10833$ trough
with $W_{\rm abs}=18.95\pm2.19$~\AA\ and three components at
$(-613,-410,-248)\,\mathrm{km\,s^{-1}}$. We fitted the FIRE spectrum using
the broad profile predicted by the stratified BLR, while allowing its
normalization and the three absorption components to vary. The model gives
$R_{50}=0.26$~pc, $\log U_{50}=-1.21$, and
$N(\mathrm{He\,I^*})=1.5\times10^{15}\,\mathrm{cm^{-2}}$. The relatively
large radius reflects the recombination origin and long lifetime of the
metastable $2\,^3{\rm S}$ level, not a low excitation energy. Each absorption
component was described by
%---------------------------------------------------------------------
\begin{equation}
 T_j(v)=1-C_{{\rm los},j}\{1-\exp[-\tau_{0,j}
 \exp(-\frac{[v-v_j]^2}{b_j^2})]\},
 \label{eq:hei_transmission}
\end{equation}
%---------------------------------------------------------------------
where $\tau_{0,j}$ follows from the \textsc{Cloudy} column and fitted $b_j$.
A three-component model is favored over two by
$\Delta\mathrm{BIC}=115$, where BIC is the Bayesian information criterion,
and gives
%---------------------------------------------------------------------
\begin{equation}
 (C_{{\rm los},1},C_{{\rm los},2},C_{{\rm los},3})=(0.41,0.67,0.67); \quad
 \chi_\nu^2=1.6.
 \label{eq:hei_abs_fit}
\end{equation}
%---------------------------------------------------------------------
All three troughs have $\tau_{0,j}\gg1$, fitted centroids
$(-514,-440,-154)\,\mathrm{km\,s^{-1}}$, and intrinsic optical-depth FWHMs
$(199,187,202)\,\mathrm{km\,s^{-1}}$; their depths therefore measure
projected covering rather than column.
The broad emission has
$\mathrm{FWHM}=1080\,\mathrm{km\,s^{-1}}$ and
$W_{\rm em}=57$~\AA, below the 210~\AA\ predicted by the unscaled cloud
calculation; line transfer and occultation can readily alter this absolute
normalization. Differences from the published component velocities reflect
the simultaneous refitting of the emission and absorption decomposition.

The two fastest He~{\sc i}$^*$ components lack Balmer counterparts and thus
cannot contain the full $10^{23}\,\mathrm{cm^{-2}}$ column of a representative
BLR cloud. Stopping models at
$\log(N_{\rm H}/\mathrm{cm^{-2}})=21.3$--21.5 saturates He~{\sc i}$^*$ while
leaving the Balmer lines nearly transparent, identifying the fast absorption
with radiation-accelerated, low-column BLR skins rather than additional dense
foreground screens.

\begin{figure}[t]
 \centering
 \includegraphics[width=0.95\columnwidth]{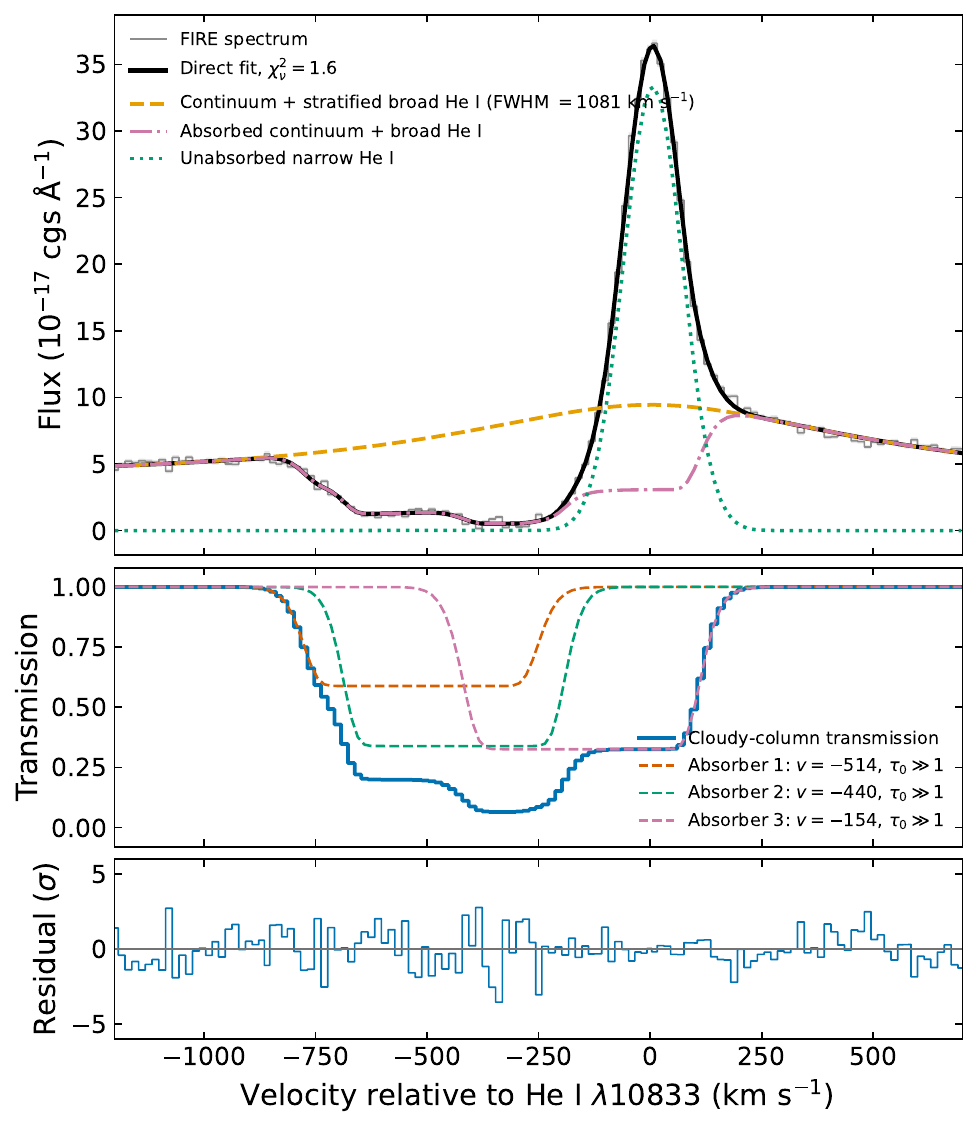}
 \caption{He~{\sc i} $\lambda10833$ in the FIRE spectrum
 \citep{Lin2026}. The upper panel shows the data and line-profile model. The
 middle panel isolates the three saturated absorption components and their
 combined transmission; the lower panel gives the residuals.}
 \label{fig:hei10830_abs}
\end{figure}

\subsection{Ca~{\sc ii} H, K, and Mg~{\sc i} absorption predictions}
\label{app:mg_absorption}

The warm CaT layer should also absorb Ca~{\sc ii} H and K. After normalizing
the prediction to CaT and accounting for dilution by starlight, we obtain
$W_{\rm Ca\,H}\simeq1.8$~\AA\ and $W_{\rm Ca\,K}\simeq1.9$~\AA. A direct
fit gives $W_{\rm Ca\,K}=1.5\pm0.3$~\AA\ and a centroid consistent with the
predicted $-65\,\mathrm{km\,s^{-1}}$ within one resolution element. This is
an encouraging independent check. Ca~{\sc ii} H is less useful because it is
blended with [Ne~{\sc iii}]~$\lambda3969$ and H$\epsilon$.

Neutral magnesium provides a complementary temperature test.
\citet{Ji2025z01} identify Mg~{\sc i}~b absorption, while
\citet{Lin2026} list possible Mg~{\sc i} absorption blended with Fe~{\sc i}.
The ground-state $\lambda2853$ line mainly traces the neutral fraction and
shielding, whereas the excited Mg~{\sc i}~b triplet is sensitive to gas
temperature. The warm layer populates the triplet sufficiently to give
$W_{\rm Mg\,b}\simeq1.0$~\AA, including the $7\%$ stellar-continuum
dilution near 5170~\AA. Mg~{\sc i}~$\lambda2853$ is saturated in the model,
but unabsorbed stellar light supplies $76\%$ of the total continuum at that
wavelength, diluting the prediction to $W_{\rm Mg\,2853}\simeq0.45$~\AA.
The published Mg identifications are blended and have no measured EWs. They
are therefore consistent with, but do not yet confirm, the warm layer. A
Mg~{\sc i}~b profile matching CaT would be a direct test against a stellar
origin.

\subsection{High-order Balmer test of the two absorbers}
\label{app:high_order_balmer}

The warm CaT layer contains enough hydrogen in the $n=2$ state to absorb the
Balmer series. Its predicted blueshifted absorption can be difficult to see
in H$\beta$--H$\delta$, where it overlaps strong unresolved emission. The
higher-order lines are cleaner, so we fitted H9--H12 simultaneously; H8 was
excluded because it blends with He~{\sc i}~$\lambda3889$. We tested two
components: the established red absorber at
$+155\,\mathrm{km\,s^{-1}}$, consistent within the resolution with the
$+142\,\mathrm{km\,s^{-1}}$ lower-order solution, and a possible warm-layer absorber at the CaT
velocity of $-66\,\mathrm{km\,s^{-1}}$. Their velocity widths were fixed by
the lower-order Balmer and CaT fits and convolved with the MODS-B resolution.
The model also includes local emission and stellar Balmer absorption, so
neither is mistaken for a nuclear trough.

The red component is clearly required: it improves $\chi^2$ by 23 and gives
$N({\rm H},n=2)\simeq(2$--$5)\times10^{15}\,\mathrm{cm^{-2}}$. The evidence
for the blueshifted warm-layer component is weak ($\Delta\chi^2=2$). Its
preferred Balmer covering is approximately 0.2, but the CaT covering of 0.35
is also allowed. Thus H9--H12 confirm the red absorber, while the Balmer
covering of the warm CaT gas remains uncertain.

\begin{figure}[!h]
\centering
\includegraphics[width=0.99\columnwidth]{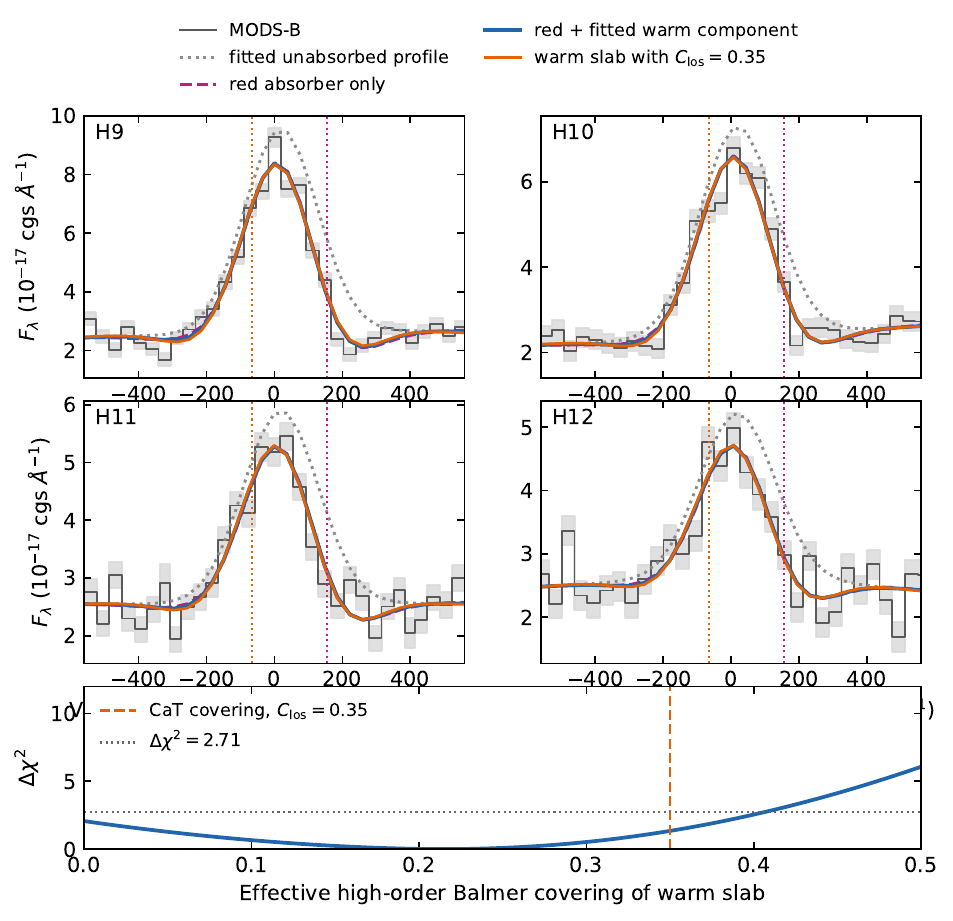}
\caption{Joint H9--H12 test. Gray profiles and bands show the MODS-B data and
$1\sigma$ uncertainties. Dashed magenta curves include only the red absorber;
blue curves add a freely fitted warm-layer absorber at the CaT velocity; and
orange curves fix its covering to the CaT value of 0.35. Vertical dotted
lines mark the two absorber velocities. The lower panel shows that the data
weakly prefer a smaller warm-layer covering but also allow 0.35.}
\label{fig:high_order_balmer_test}
\end{figure}

\subsection{Warm CaT-layer heating diagnostics}
\label{app:cat_heating}

The preferred $0.1\,Z_\odot$ model uses a non-radiative heating rate
$H_{\rm nr}=1.2\times10^{-5}\,\mathrm{erg\,cm^{-3}\,s^{-1}}$; H$^-$ then
supplies at most $5\%$ locally, and free--free
heating is subdominant. Because projected covering does not fix global sky
covering $C_{\rm warm}$, integration gives
$L_{\rm nr}\simeq5\times10^{45}C_{\rm warm}\,\mathrm{erg\,s^{-1}}$, or
$10^{44}\,\mathrm{erg\,s^{-1}}$ for $C_{\rm warm}=0.02$. The orbital reservoir
would last only two months, requiring sustained shocks or collisions,
turbulence, energetic particles, or magnetic transport
\citep{CollinJoly2000,VeronCetty2004,KrolikBegelman1988,Vollmer2004,Vollmer2022,Ferland2009}.
A solar-metallicity calculation with unit projected covering matches CaT at
$H_{\rm nr}\simeq(2$--$3)\times10^{-8}\,\mathrm{erg\,cm^{-3}\,s^{-1}}$,
2.7 dex below the preferred value, illustrating the abundance and covering
dependence. Tentative Fe/Ti identifications lack secure EWs and do not
constrain metallicity \citep{Ji2025z01,Lin2026}.

\onecolumn
\section{Kinematic and observational summary}
\label{app:scorecard}
\begin{table}[h!]
\caption{Velocity ladder of the emitting and absorbing phases.}
\label{tab:velocity_ladder}
\centering
\small
\begin{tabular}{lccccc}
\hline\hline
Tracer or phase & $\Delta v$ & $r$ & $v_{\rm esc}$ &
$|\Delta v|/v_{\rm esc}$ & Interpretation \\
 & ($\mathrm{km\,s^{-1}}$) & (pc) & ($\mathrm{km\,s^{-1}}$) & & \\
\hline
He~{\sc i}$^*$ absorption skins & $-514,-440,-154$ & 0.26 & 910 & 0.17--0.56 & ionized outflow \\
H$\alpha$ absorption & $-130$ & $\sim1.9$ & 340 & 0.38 & outflowing outer complex \\
Na~{\sc i}, K~{\sc i} absorption & $-90$ & 6.6 & 180 & 0.50 & cool dusty gas (possibly outflowing) \\
CaT absorption & $-66$ & $\sim1.9$ & 340 & 0.19 & locally heated interface \\
H$\alpha$ emission pile-up & $+17$ & $\sim1.9$ & 340 & 0.05 & integrated outer complex \\
H$\beta$--H$\delta$ absorption & $+140$ to $+170$ & $\sim1.9$ & 340 & 0.41--0.50 & infalling outer complex \\
\hline
\end{tabular}
\tablefoot{Velocities are projected line-of-sight values
relative to the adopted systemic redshift, $z=0.1006$.}
\end{table}

\begin{table}[h!]
\caption{Summary of observational tests of the preferred model.}
\label{tab:scorecard}
\centering
\footnotesize
\setlength{\tabcolsep}{4pt}
\begin{tabular}{p{0.20\textwidth}p{0.60\textwidth}p{0.12\textwidth}}
\hline\hline
Observable & Model comparison & Assessment \\
\hline
UV--near-IR continuum & Disk plus young stars gives
$\chi_\nu^2=0.9$ without arbitrary disk renormalization. & \scoreGood \\
Warm infrared emission & Fixed continuum plus warm dust gives
$\chi_\nu^2=0.2$; the far-IR peak is not observed. & \scoreGood \\
X-ray weakness & Line-of-sight absorption reduces the predicted thermal
$\nu L_\nu(2\,\mathrm{keV})$ to $(3.4$--$4.1)\times10^{40}\,
\mathrm{erg\,s^{-1}}$, below the \textit{Chandra} limit. & \scorePartial \\
Physical parameters & The black-hole mass is consistent with dust-corrected
virial estimates. The SED value $i=73\degr$ is influenced substantially by
the inclination prior, while $\dot m$ and $R_V$ reach grid boundaries.
& \scorePartial \\
Broad H$\alpha$ emission & The measured
$W_{\rm H\alpha}=357\pm18$~\AA\ fixes
$C_{\rm BLR}=0.02$. & \scoreCalibrated \\
Broad H$\alpha$ emission profile and wings & The stratified model reproduces the full
H$\alpha$ profile, including its extended wings, with
$\chi_\nu^2=1.1$. & \scoreGood \\
Balmer decrement & Predicted H$\alpha$/H$\beta=17.6$; observed
$16.9\pm1.8$. & \scoreExcellent \\
Balmer stratification & Predicted
$R_{50}({\rm H\alpha})/R_{50}({\rm H\beta})=1.6$ and
$R_{50}({\rm H\alpha})/R_{50}({\rm H\delta})=1.6$; the Cloudy-normalized broad
H$\beta$--H$\delta$ emission profiles reproduce the observed wings. & \scoreGood \\
Balmer absorption & A common red H$\beta$--H$\delta$ absorber is
recovered. The modeled H$\alpha$ and H$\beta$ emitting regions have similar
projected covering at the absorber velocity, so its missing H$\alpha$
counterpart requires velocity-coherent H$\alpha$ refill.
& \scorePartial \\
Outer-BLR pile-up & The continuous radial model reproduces the H$\alpha$
core and shoulders without a separate Gaussian. A Pa$\alpha$ emission counterpart is
favored, whereas the higher-Balmer emission contribution requires selective transfer.
& \scorePartial \\
He~{\sc i} $\lambda10833$ absorption and emission & Low-column ionized BLR
skins reproduce the saturated absorption-trough kinematics. The full stratified
BLR reproduces the broad-emission FWHM after its normalization is allowed to vary.
& \scorePartial \\
Paschen decrement and normalization & Pa$\alpha$/Pa$\gamma\simeq5.2$ versus $7.46\pm0.33$; the
baseline inner-BLR Pa$\alpha$ flux is overpredicted by a factor of order three.
& \scoreFail \\
High-excitation lines & Broad He~{\sc ii} $\lambda4686$ has predicted
$W_{\rm b}\simeq2.3$~\AA, qualitatively consistent with the absence of a secure
broad detection; quantitative EW limits are unavailable. & \scorePartial \\
Na~{\sc i} and K~{\sc i} absorption & Nested saturated Na~{\sc i}
coverage gives $6.0$~\AA\ versus $6.6\pm0.4$~\AA; the cool and warm
K~{\sc i} contributions sum to $3.2$~\AA\ versus $3.1\pm0.1$~\AA.
& \scoreGood \\
Ca~{\sc ii} triplet absorption & A warm boundary layer with
non-radiative heating matches CaT absorption. The CaT
and H$\alpha$ absorbers have compatible centroids within one resolution
element but prefer different covering fractions; H9--H12 nevertheless allow the CaT covering fraction
(Appendix~\ref{app:high_order_balmer}).
& \scorePartial \\
Continuum and broad H$\alpha$ polarization & At fixed continuum parameters, the model gives
$p_{\rm cont}=1.62\%$, $p_{\rm H\alpha}=0.80\%$, and
$\Delta\vartheta=-52\degr$, versus $1.53\%$, $0.58$--$0.84\%$, and
$-48\degr$ observed. & \scoreExcellent \\
\hline
\end{tabular}
\tablefoot{\scoreExcellent\ denotes excellent quantitative agreement;
\scoreGood\ denotes a good fit or consistency with an upper limit or external
estimate; \scorePartial\ denotes partial agreement with a remaining discrepancy;
\scoreCalibrated\ identifies a datum used for calibration; and \scoreFail\ denotes
an observable that is not reproduced.}
\end{table}

\end{appendix}

\end{document}